\documentclass[12pt]{article}
\usepackage{amsmath}
\usepackage{graphicx}
\usepackage{enumerate}
\usepackage{natbib}
\usepackage{url} 
\usepackage{enumitem}
\usepackage{bbm}
\usepackage{subfig}
\usepackage{multirow}
\usepackage{amssymb,amsmath}
\usepackage{mathrsfs}
\usepackage{epsfig,morefloats}
\usepackage{amsfonts}
\usepackage{amsthm}
\usepackage{lscape}
\usepackage{rotating}
\usepackage{float}
\usepackage{caption,hyperref}
\usepackage{color}
\usepackage{tabularx}
\usepackage{times}
\usepackage{lineno}
\usepackage[ruled,vlined]{algorithm2e}
\usepackage{setspace}
\newcommand{\blind}{1}
\def\be{\begin{equation}}
\def\ee{\end{equation}}
\def\bea{\begin{eqnarray}}
\def\eea{\end{eqnarray}}
\def\nn{\nonumber}

\def\br{\mathrm{BR}}
\def\R{\mathbb{R}}

\def\pr{\mathbb{P}}
\definecolor{hyw}{RGB}{153,000,000}

\definecolor{yrz}{RGB}{000,000,200}

\def\cG{\mathcal{G}}

\def\cI{\mathcal{I}}

\def\cF{\mathcal{F}}

\def\E{\mathbb{E}}

\def\I{\mathbb{I}}

\newcommand{\V}{\rm Var}

\newtheorem{assumption}{Assumption}
\newtheorem{proposition}{Proposition}

\newtheorem{theorem}{Theorem}

\newtheoremstyle{remarkstyle}
{}                    
{}                   
{\normalfont}         
{}                   
{\itshape}            
{.}                   
{ }                  
{}                   
\theoremstyle{remarkstyle}
\newtheorem{remark}{Remark}
\begin{document}

\def\spacingset#1{\renewcommand{\baselinestretch}
{#1}\small\normalsize} \spacingset{1}

\if1\blind
{
\title{\bf Collaborative Indirect Treatment Comparisons with Multiple Distributed Single-arm Trials}
\author{Yuru Zhu
\hspace{.2cm}\\
Department of Biostatistics, Epidemiology, and Informatics, University of Pennsylvania\\
Huiyuan Wang\hspace{.2cm}\\
Department of Biostatistics, Epidemiology, and Informatics, University of Pennsylvania\\
Haitao Chu \hspace{.2cm}\\
Global Biometrics and Data Management, Pfizer Inc.\\
Division of Biostatistics and Health Data Science, University of Minnesota Twin Cities\\
Yumou Qiu \hspace{.2cm}\\
School of Mathematical Sciences, Peking University\\
and \\
Yong Chen\thanks{corresponding author} \\
Department of Biostatistics, Epidemiology, and Informatics, University of Pennsylvania\\
}
\maketitle
} \fi

\if0\blind
{
\bigskip
\bigskip
\bigskip
\begin{center}
{\LARGE\bf Collaborative Indirect Treatment Comparisons with Multiple Distributed Single-arm Trials}
\end{center}
\medskip
} \fi

\bigskip
\begin{abstract}
When randomized controlled trials are impractical or unethical to simultaneously compare multiple treatments, indirect treatment comparisons using single-arm trials offer valuable evidence for health technology assessments, especially for rare diseases and early-phase drug development. In practice, each sponsor conducts  a single-arm trial on its own drug with restricted data-sharing and targets effects in its trial population, which can lead to unfair comparisons.
This motivates methods for fair treatment comparisons across a range of target populations in distributed networks of single-arm trials sharing only aggregated data. Existing federated methods, which assume at least one site contains all treatments and allow pooling of treatment groups within the same site, cannot address this problem. We propose a novel distributed augmented calibration weighting (DAC) method to simultaneously estimate the pairwise average treatment effects (ATEs) across all trial population combinations in a distributed network of multiple single-arm trials. Using two communication rounds, DAC estimators balance covariates via calibration weighting, incorporate flexible nuisance parameter estimation, achieve doubly robust consistency, and yield results identical to pooled-data analysis.  
When nuisance parameters are estimated parametrically, DAC estimators are enhanced to achieve doubly robust inference with minimal squared first-order asymptotic bias. Simulations and a real-data application show good performance.
\end{abstract} 

\vfill

\newpage
\spacingset{1.9} 
\section{Introduction}
\label{sec:intro}
Generating evidence for relative effectiveness of multiple interventions is crucial in modern healthcare, with applications in drug development, public health policy evaluation, and precision medicine \citep{miksad2019small, dagenais2022use, william2023the}. 
While randomized controlled trials that jointly evaluate all treatments remain the gold standard, they are often infeasible in practice due to strict eligibility criteria, the growing number of treatment options, or ethical concerns. For example, guidelines developed by the U.S. Food and Drug Administration \citeyearpar{FDA2001guidance} discourage placebo controls when an available treatment is known to prevent serious harm, such as death or irreversible morbidity.  
In such cases, single-arm trials, where all participants receive a predetermined treatment, offer an alternative. These trials are widely used for rare diseases and early-phase drug development, and have even supported regulatory approvals \citep{vickers2007setting,seymour2010design,hatswell2016regulatory}. 

Because single-arm trials lack concurrent comparators, auxiliary data are often used to impute the ``missing'' comparator group, enabling indirect treatment comparisons (ITCs) in an ``unanchored'' manner \citep{phillippo2016nice,thorlund2020synthetic, jahanshahi2021use}. 
ITCs play a critical role in health technology assessment (HTA) and joint clinical assessment (JCA) in pharmaceutical research, where timely and fair evaluation of new therapies is important. To support it, the HTA Regulation \citep{eu2021hta,eu2024jca1,eu2024jca2} in the European Union mandates consistent, transparent, and comparative evaluation of clinical evidence. 
In drug development, each sponsor typically evaluates only its own investigational drug, while access to competitors’
data is restricted by commercial confidentiality or prohibitive licensing costs.
Sponsors, assessors and regulators often conduct ITCs by emulating comparator groups using summary statistics \citep{phillippo2016nice,patel2021use}, leading to distributed networks of single-arm trials, particularly in rare diseases where placebo controls are infeasible. Furthermore, equitable assessment requires treatment comparison on a common target population, but sponsors usually prioritize their own trial population, yielding misaligned evaluations. 

Distributed networks permit sharing aggregated data but face challenges, including efficiency loss due to distributed inference and heterogeneous covariate distributions across sites, and communication burden \citep{gu2023distributed}.  
Most existing federated causal inference settings consider that at least one site contains all treatments of interest \citep{Xiong2023Federated,Hu2024Collaborative,Han2025Federated,Guo2025Robust}. In such cases, the causal effect can
be identified using data from that site alone, and federation primarily improves efficiency. Moreover, these methods estimate the effect for
one prespecified target population per run, requiring repeated executions for multiple populations and thus increasing communication costs.

These challenges motivate developing methods that (i) support ITCs within distributed networks of multiple single-arm trials, and (ii) enable simultaneously comparing multiple treatments across a range of target
populations, including all possible combinations of single-arm trial populations, while respecting data-sharing constraints and minimizing communication burden with no efficiency loss. 
To this end, we propose a novel Distributed Augmented Calibration weighting (DAC) method for ITCs in a distributed network of multiple single-arm trials, where each site retains its data locally and shares aggregated data with a coordinating center. 

Our contributions are twofold. First, we introduce a new federated causal inference setting where each site conducts a single-arm trial for a unique treatment and individual-level data cannot be shared across treatment groups. 
This setting, unlike existing work that allows pooling of treatment groups within the same site, imposes stricter data-sharing constraints, making federation essential for identification.
It reflects practical scenarios in HTA submissions for comparative effectiveness evidence of new drugs and naturally extends to federated causal inference settings where each site evaluates a subset of treatments. 

Second, we develop the DAC estimation method to address the unique challenges of this setting.
DAC leverages calibration weighting \citep{Deville1992Calibration} with summary statistics to balance covariates. DAC estimators accommodate flexible parametric or nonparametric nuisance estimation, and achieve doubly robust consistency.
Under parametric models, our bias-reduced DAC provides the first adaptation of the bias-reduction technique \citep{vermeulen2015bias} to the distributed network of single-arm trials with a tailored algorithm, enabling doubly robust inference with minimal squared first-order asymptotic bias under strict data-sharing constraints.
Crucially, our DAC estimator exactly recovers the pooled augmented calibration weighting estimator using all individual-level data, ensuring no information loss.
With only two communication rounds of aggregated data, DAC addresses heterogeneous covariate distributions, protects privacy, avoids information loss, and delivers valid inference for all pairwise average treatment effects (ATEs) across all possible target populations formed from site combinations. 

\subsection{Related Work}
The basis of using the internal single-arm trial data with its external auxiliary data to conduct indirect comparison is the assumption that links mechanisms of generating potential outcomes across data sources, allowing the generalization of potential outcomes from a study sample to a target population.  
When individual-level auxiliary data on alternative treatments are available, the distributional exchangeability assumption, which ensures counterfactual outcomes across sources have the same conditional distribution given observed confounders, allows combining all data as a multi-treatment study. Causal methods such as stratification, matching, modeling, and weighting can then estimate target ATEs \citep{Lee2018Efficient,Li2020Target,shi2023data,Wang2024comparison}.

When only limited summary statistics are available, estimation becomes more restrictive. Meta-analysis methods rely on an assumption that the expected counterfactual outcome has  same value or distribution across sources, but may yield biased estimates when there are systematic differences between the single-arm trial and auxiliary data populations  \citep{schmidli2014robust,woolacott2017methodological,weber2018predicting}. 
The unachored matching-adjusted indirect comparison (MAIC) approach \citep{Signorovitch2010Comparative} combines the internal single-arm trial data with some fixed external summary statistics of the alternative treatment  via calibration weighting \citep{phillippo2018methods, Cheng2020statistical}. MAIC addresses covariate heterogeneity but only identifies the treatment effect in the external data population and depends on correctly specifying the selection model for inclusion in the internal single-arm trial, as poorly chosen or irelevant variables can introduce bias 
\citep{Deville1992Calibration,Chan2015Globally,Zubizarreta2015stable}.
However, summary statistics for calibration are often limited in practice, especially when comparing an increasing number of treatments simultaneously under complex selection mechanisms. 

In contrast, our DAC approach requires two communication rounds to transfer specific aggregated data, balancing patient privacy with the limitations of fixed summary statistics. 
DAC handles heterogeneous covariate distributions, supports simultaneous pairwise comparisons across a range of target populations and achieves the same results as pooled individual-level analysis.

\subsection{Overview of the Paper}
Section \ref{sec:preliminary} outlines the problem setup and identification of the estimands in a distributed network. 
Section \ref{sec:DRAWmethod} introduces the general DAC estimation approach and its algorithm with nonparametric methods for estimating working models. Section \ref{sec:BRDRAW} provides the bias-reduced DAC estimation for doubly robust inference under parametric working models. Section \ref{sec:theory} illustrates the asymptotic properties of the proposed DAC estimators.
Sections \ref{sec:simulation} and \ref{sec:empirical} show performance of the DAC estimator in simulations and empirical data analysis. Section \ref{sec:conclusion} concludes with a discussion.

\section{Preliminaries}
\label{sec:preliminary}

We consider a distributed network consisting of $K$ finite sites.  Let the categorical random variable $D\in\{1,2,\dots,K\}$ index the site and $X$ denote a vector of pre-treatment covariates that explain treatment assignments, such as demographic characteristics and inclusion/exclusion criteria. 
After selecting eligible patients, each site conducts a single-arm trial and collects the outcome $Y$. 
At site $k$, we observe $n_k$ independent and identically distributed samples $\{(D_i, Y_i,X_i)\}_{D_i=k}$ drawn from its underlying single-arm trial population whose distribution is denoted as $P_k$. Let $f_{X\mid D=k} (x)$ be the density of covariates $X$ at site $k$. Denote the total sample size of all sites as $N=\sum_{d=1}^K n_d$ with the limiting proportion $\pr(D=k):=\lim_{N\to \infty} n_k/N>0$ for each site $k\in\{1,2,\dots,K\}$. 
For better illustration, we focus on the case where no treatment assignments are repeated across sites, thus $D$ can also represent the treatment assignment. Extensions to cases of repeated treatments at different sites or multiple treatments assigned within a local site are discussed in Section S6 of the Supplementary Material (SM).

This paper aims to develop an estimation procedure that, in a single implementation, enables simultaneously comparing the effectiveness of various treatments across different subpopulations. Specifically, we address the following questions for each site:
\begin{enumerate}[label=(Q.\arabic*)]
\item \label{q1} How would outcomes change if its local patients received a different treatment?
\item \label{q2} What would a relative effect be if treatments were compared in another patient population?
\end{enumerate}
We approach these questions using the potential outcome framework \citep{rubin1974estimating}. 
Let $Y(d)$ be the potential outcome observed if the subject were assigned treatment $d$. 
The causal estimand 
\begin{equation}\label{eq:estimand}
\tau_{(k,k')\mid\cI}=\E\{Y(k')\mid D\in\cI\}-\E\{Y(k)\mid D\in\cI\}, \quad k, k' =1, \cdots, K, k \neq k',
\end{equation}
represents the comparative effectiveness of treatment $k'$ over $k$ in the subpopulation of patients in the sites indexed by a subset $\cI\subseteq\{1,\dots,K\}$, whose covariate density is $\{\sum_{d\in\cI} \pr(D=d)\}^{-1}\sum_{d\in\cI} \pr(D=d) f_{X\mid D=d}(x)$. The cases $\cI=\{k\}$ and $\cI\ne\{k\}$ correspond to Questions \ref{q1} and \ref{q2}, respectively. Rather than focusing on one pre-specified target population, we consider all subpopulations corresponding to site subsets $\cI \subseteq \{1, \cdots, K\}$. Since the estimand in \eqref{eq:estimand} is a difference in expectations of two potential outcomes, it suffices to identify the expected potential outcome $\mu_{k\mid\cI}\equiv\E\{Y(k)\mid D\in\cI\}$ in the target population. 
We introduce the following assumptions to establish the identifiability of $\mu_{k\mid\cI}$  under the distributed data structure.

Following the stable unit treatment value assumption that there is no interference between units and no multiple versions of treatment, we make the consistency assumption:
\begin{assumption}[Consistency]\label{as:1}
The observed outcome is $Y=\sum_{d=1}^KY(d)I(D=d)$.
\end{assumption} 

By Assumption~\ref{as:1}, the observed outcome at site $k$ is the potential outcome under treatment $k$.

\begin{assumption}[Exchangeability]\label{as:2}
The conditional distributions of potential outcomes given covariates are identical across sites: $(Y(1),\dots,Y(K)) \mid (X, D=j) \overset{d}{=} (Y(1),\dots,Y(K)) \mid (X, D=k)$ for any $j, k \in \{1, \cdots, K\}$ with $j \neq k$.
\end{assumption}
This assumption posits that treatments operate via the same mechanism across different single-arm trial populations given observed covariates, ensuring no unmeasured confounding between sites (treatments) and potential outcomes \citep{Stuart2011Use,shi2023data}. If local data were pooled, Assumption~\ref{as:2} implies that the $K$ single-arm trials would collectively resemble a hypothetical $K$-arm randomized trial.  
For single-arm trials investigating different treatments for the same disease, 
aligned eligibility criteria within similar time periods make this assumption reasonable in our context \citep{gray2020framework}. 
Although the exchangeability assumption is generally not testable, Section S6.3 of the SM discusses its assessment using negative control outcomes and provides practical guidance for mitigating bias when this assumption may be violated.
To ensure comparability across sites, we impose the following positivity assumption. 
\begin{assumption}[Positivity]\label{as:3} 
For a positive constant $\varepsilon$, 
$\min_{j, k=1,\dots,K} \frac{f_{X\mid D=j} (x)}{f_{X\mid D=k} (x)} \geq \varepsilon$
almost surely.
\end{assumption}
Assumption \ref{as:3} is plausible, as eligibility criteria for single-arm trials conducted in close time periods are often similar, leading to identical supports for covariate distributions across different single-arm trials. Let $Z=(D,Y,X^T)^T$ and $P$ denote the distribution of $Z$ for the overall population combining all single-arm trial populations, i.e., $P=\sum_{d=1}^{K} \pr(D=d) \times P_d$.
Let $m_k(X)\equiv\E(Y\mid X,D=k)$ be the conditional mean function  of outcome $Y$ given $X$ at the local site $k$ and $w_{jk}(x)\equiv \{f_{X\mid D=j} (x) \pr(D=j)\}/\{f_{X\mid D=k} (x) \pr(D=k)\}$  be a weighting function for  $j,k\in \{1,\dots,K\}$.
Then, the identification is presented as follows.
\begin{proposition}\label{prop:iden}
Under Assumptions~\ref{as:1}–\ref{as:3}, for any $k\in \{1,\dots,K\}$ and $\cI\subseteq\{1,\dots,K\}$, the expected potential outcome $\mu_{k\mid\cI}$ can be identified in two  forms:
\begin{align}
    \mu_{k\mid\cI} & =\E_P\{\frac{m_k(X)I(D\in\cI)}{\pr(D\in\cI)}\}. \label{eq:or}\\
    \mu_{k\mid\cI} &=\E_P\biggl\{\frac{YI(D=k)}{\pr(D\in\cI)}\times\sum_{j\in\cI}w_{jk}(X)\biggr\}. \label{eq:ps}
\end{align}
\end{proposition}

Proposition \ref{prop:iden} leads to two plug-in estimators for $\mu_{k\mid\cI}$. The first is the outcome regression (OR) estimator $\hat\mu_{k\mid\cI}^{\text{OR}}=\{\sum_{i=1}^N\hat{m}_k(X_i)I(D_i\in\cI)\}/\{\sum_{j\in\cI}n_j\}$, which averages predictions of the estimated conditional mean function $\hat{m}_k(X)$ across patients in sites indexed by $\cI$.
The second is the propensity score weighting (PS) estimator $\hat\mu_{k\mid\cI}^{\text{PS}}=\{\sum_{i=1}^NY_iI(D_i=k)\sum_{j\in\cI}\hat{w}_{jk}(X_i)\}/\{\sum_{j\in\cI}n_j\}$, which plugs the estimated weighting function $\hat{w}_{jk}(X)$ in \eqref{eq:ps}. 
These methods are widely used for causal effect estimation \citep{shi2023data}. 
However, consistency of OR or PS estimators relies on correct specifications of regression models or propensity score models, respectively \citep{Lunceford2004strat,funk2011doubly}. 
Moreover, estimating weighting functions in distributed settings without sharing individual-level data poses challenges.
To address these issues, the next two sections introduce our Distributed Augmented Calibration weighting (DAC) method. DAC leverages the two forms of identification in Proposition \ref{prop:iden} to provide robust estimators for simultaneously comparing treatment effects across various target populations, requiring only aggregated data shared within minimal communication rounds.

\section{Distributed Augmented Calibration Method}
\label{sec:DRAWmethod}
In this section, we propose the general DAC approach for doubly robust estimation of $\tau_{(k,k')\mid\cI}$ for any $k, k' \in \{1, \cdots, K\}, k \neq k'$ and $\cI \subseteq \{1, \cdots, K\}$, which can incorporate nonparametric methods for estimating outcome regression models and weighting functions.
We begin by deriving the efficient influence functions (EIFs) for target estimands. The EIF for $\mu_{k \mid \mathcal{I}}$ is
\begin{equation*} 
\varphi_{k\mid\cI}\bigl(Z;P\bigr)=\frac{1}{\pr(D\in\cI)}\biggl[I(D\in\cI)\bigl\{m_k(X)-\mu_{k\mid\cI}\bigr\}+I(D=k)\bigl\{Y-m_k(X)\bigr\}\sum_{j\in\cI}w_{jk}(X)\biggr], 
\end{equation*}  
which corrects the first-order bias of plug-in estimators (see Section~2.2.5 in \citealt{chernozhukov2018double}).
Similary, the EIF for $\tau_{(k,k')\mid\cI}$ is $\phi_{(k,k')\mid\cI}\bigl(Z;P\bigr)=\varphi_{k'\mid\cI}\bigl(Z;P\bigr)-\varphi_{k\mid\cI}\bigl(Z;P\bigr)$.
The derivation of these EIFs is detailed in Theorem \ref{pro:2} in Section \ref{sec:theory}. 

From above EIFs, our DAC requires estimating two working models: the outcome model $m_k(X)$ and weighting function $w_{jk}(X)$. While $m_k(X)$ can be estimated locally at site $k$, estimating $w_{jk}(X)$ without accessing individual-level data is challenging due to data-sharing restrictions and population heterogeneity in distributed settings. Instead of directly modeling the sampling propensity score, we adopt the calibration weighting (CW) approach \citep{Deville1992Calibration}, which uses only summary-level statistics to balance covariates and avoids extreme weights \citep{Qin2007empirical,Hainmueller2012entropy,Zubizarreta2015stable}.
It is based on the identity 
$\E[ I(D=k)w_{jk}(X)g(X)/ \pr(D=j)]  
=  \E\{g(X) \mid D=j\}$,  
where  ${g}(X)$ is a vector of functions of $X$ (e.g., moments of $X$) to be calibrated. Its empirical analog leads to the calibration constraints $\sum_{D_i=k}w_{jk}(X_i)/n_j g(X_i) = \bar{{g}}_j(X)$, where $\bar{{g}}_j(X)=\sum_{D_i=j} {g}(X_i)/n_j$ is the sample mean of ${g}(X)$ at site $j$, suggesting aligning the weighted mean of $g(X)$ from site $k$ with the sample mean of $g(X)$ at site $j$ to achieve covariate balance between the two sites. Weights are estimated by minimizing a loss function subject to these balancing constraints:
\begin{equation}
\begin{split}
&\min_{\{w_{jk}(X_i), D_i=k\}}\sum_{D_i=k} {\rm d}(w_{jk}(X_i)/n_j, 1/n_j) \\ 
&\text{ subject to } w_{jk}(X_i) \geq 0,  \sum_{D_i=k} w_{jk}(X_i)/n_j =1 \text{ and } \sum_{D_i=k} w_{jk}(X_i)/n_j {g}(X_i) = \bar{{g}}_j(X).
\end{split}
\label{eq:cw}
\end{equation}
Here, we use  the negative entropy loss ${\rm d}(w_{jk}(X_i)/n_j, 1/n_j)=w_{jk}(X_i)\log\{w_{jk}(X_i)\}/n_j$, which measures the distance between the weight $w_{jk}(X_i)/n_j$  and the uniform weight $1/n_j$ to minimize the variability from heterogeneous weights, corresponding to the entropy balancing method \citep{Hainmueller2012entropy}. 
By introducing Lagrange multiplier $\gamma_{jk}$, 
the estimated calibration weight is 
\begin{equation}
\hat{w}_{jk}(X_i)/n_j = \exp\{\hat{\gamma}_{jk}^{T} {g}(X_i)\}/ \sum_{D_i=k} \exp\{\hat{\gamma}_{jk}^{T} {g}(X_i)\} \text{ for } D_i=k,
\label{eq:weights}
\end{equation}
where $\hat{\gamma}_{jk}$ solves the estimating equation 
\begin{equation}
\sum_{D_i=k}\exp\{\gamma_{jk}^{T} {g}(X_i)\}\{{g}(X_i) - \bar{{g}}_j(X)\} = 0.
\label{eq:optimize}
\end{equation}
It is the Lagrangian dual of optimization problem (\ref{eq:cw}) under the negative entropy loss, which implies a log-linear model $\log\{w_{jk}(X)\}= \alpha_{jk} +\gamma_{jk}^T{g}(X)$  \citep{Cheng2020statistical}.

The DAC estimator based on the estimated outcome model $\hat{m}_k$ and weighting function $\hat{w}_{jk}$ is
\begin{eqnarray}
&\hat\mu_{k\mid\cI} = \sum_{j\in \mathcal{I}} \frac{n_j}{\sum_{j\in \mathcal{I}}n_j} ( A^1_{jk}   + A^2_{kj} ), \label{eq:dr_mu} \\ 
&\hat{\tau}_{(k,k')\mid\cI} =\sum_{j \in \mathcal{I}} \frac{n_j}{\sum_{j\in \mathcal{I}}n_j} ( A^1_{jk'}  -
A^1_{jk}   + A^2_{k'j} - A^2_{kj}). 
\label{eq:DRmultiple}
\end{eqnarray}
where  
\begin{equation}
A^1_{jk}=\frac{1}{n_j}\sum_{i:D_i=j}\hat{m}_k(X_i), \ A^2_{kj}= \frac{1}{n_j} \sum_{i:D_i=k} \hat{w}_{jk}(X_i) \{Y_i-\hat{m}_k(X_i)\}
\label{eq:A12}
\end{equation}
are two types of aggregated data. 
The first is the sample mean of $\hat{m}_k(X)$ at local site $j$,  the second is a weighted average of residuals of $\hat{m}_k$ at local site $k$. And $\hat\mu_{k\mid\cI}$ is a weighted average of the sums of these two aggregated data across sites in $\cI$, enabling privacy-preserving estimation. 
 
\begin{remark}
Using individual-level data from all single-arm trials, an estimator by solving $\E_P\bigl\{\phi_{(k,k')\mid\cI}$ $\bigl(Z;P\bigr)\bigr\}=0$ for $\tau_{(k,k')\mid\cI}$ empirically based on respective estimators $\hat{m}_l(X)$ and $\hat{w}_{jl}(X)$ for $m_l(X)$ and $w_{jl}(X)$, $l=k, k', j \in \cI$, has an explicit form $\hat\tau_{(k,k')\mid\cI}^{\text{pool}}=\hat\mu_{k'\mid\cI}^{\text{pool}}-\hat\mu_{k\mid\cI}^{\text{pool}}$, where
\begin{equation*}
\hat\mu_{k\mid\cI}^{\text{pool}}=\frac{1}{\sum_{j\in\cI}n_j}\sum_{i=1}^N\biggl\{\hat{m}_k(X_i)I(D_i\in\cI)+I(D_i=k)\bigl\{Y_i-\hat{m}_k(X_i)\bigr\}\sum_{j\in\cI}\hat{w}_{jk}(X_i)\biggr\}.
\end{equation*}
Notably, our DAC estimator for $\tau_{(k,k')\mid\cI}$ in \eqref{eq:DRmultiple} using only aggregated data replicates $\hat\tau_{(k,k')\mid\cI}^{\text{pool}}$ using pooled data. Thus, our approach is as efficient and robust as $\hat\tau_{(k,k')\mid\cI}^{\text{pool}}$ without information loss.
\end{remark}

\begin{figure}[!ht]
\centering
\includegraphics[width=0.99\textwidth]{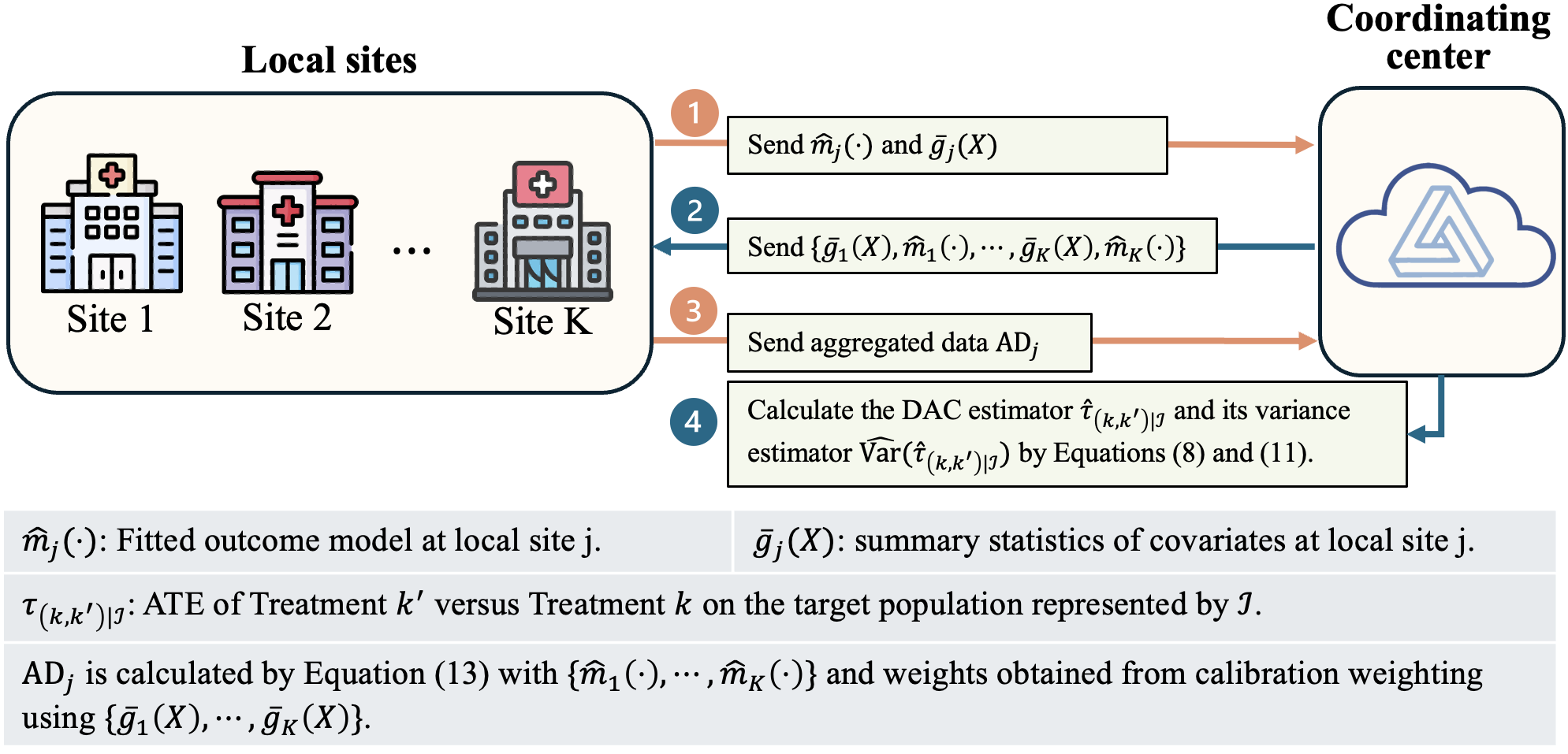}
\caption{Workflow of the algorithm on the distributed augmented calibration weighting (DAC) estimation under nonparametric working models for the collaborative
indirect comparisons in a distributed network where each site conducts a single-arm trial on a distinct treatment.
\label{fig:workflowDR} }
\end{figure}

\begin{remark}
    The DAC estimator for $\mu_{k\mid\cI}$ in \eqref{eq:dr_mu} differs from the distributed calibration weighting (DCW) estimator, $\hat\mu_{k\mid\cI}^{\text{DCW}}:=\sum_{j \in \mathcal{I}} {n_j}B^2_{kj}/(\sum_{j\in \mathcal{I}}n_j)$, where $B^2_{kj}= \sum_{i: D_i=k} \hat{w}_{jk}(X_i) Y_i/n_j$, by 
    \begin{equation}
        \sum_{j\in \mathcal{I}} \frac{n_j}{\sum_{j\in \mathcal{I}}n_j} \{\sum_{i:D_i=j}\frac{\hat{m}_k(X_i)}{n_j} - \sum_{i:D_i=k} \frac{\hat{w}_{jk}(X_i)}{n_j} \hat{m}_k(X_i)\}. \label{eq:DACCWdif}
    \end{equation}
    This term serves as the bias-correction component, ensuring that the DAC estimator which incorporates local average of $\hat{m}_k(X)$ remains consistent under a correctly specified outcome model even when the weighting function is misspecified, in contrast to the DCW estimator.
    This difference is zero if calibration constraints in \eqref{eq:cw} also balance $\hat{m}_k(X)$, making the DAC and DCW estimators equivalent when CW balances both summary statistics $g(X)$ and the fitted outcome model $\hat{m}_k(X)$. However, this requires an additional communication round to share local averages of $\hat{m}_1(X), \cdots, \hat{m}_K(X)$ at each site for obtaining $\hat{\tau}_{(k,k')\mid\cI}^{\text{DCW}}=\hat\mu_{k'\mid\cI}^{\text{DCW}}-\hat\mu_{k\mid\cI}^{\text{DCW}}$, as shown in Figure S1 in the SM. To improve communication efficiency, our DAC method combines the OR with the CW approach that balances $g(X)$ alone, requiring two rounds as shown in Figure \ref{fig:workflowDR}. 
\end{remark}

The DAC estimator in (\ref{eq:DRmultiple}) combines the OR estimator with a bias correction term and is doubly robust consistent if either $m_k(X)$ or $w_{jk}(X)$ is consistently estimated; see Theorem \ref{prop:dr_consistency} for details. The semiparametric efficiency bound for $\tau_{(k,k')\mid\cI}$ is $B_{(k,k')\mid\cI}=\E_P\bigl\{\phi^2_{(k,k')\mid\cI}\bigl(Z;P\bigr)\bigr\}$, which means no estimator can have a smaller mean squared error than $B_{(k,k')\mid\cI}$ in the local minimax sense (see Corollary~2.6 in \citealt{vander2002}).
As stated in Theorem \ref{prop:npara} in Section \ref{sec:theory}, if both outcome models and weighting functions are consistently estimated with a convergence rate of at least $N^{1/4}$, the asymptotic variance of $\hat{\tau}_{(k,k')\mid\cI}$ attains $B_{(k,k')\mid\cI}$, yielding the variance estimation:
\begin{equation}
\begin{split}
\widehat{\V}(\hat{\tau}_{(k,k')\mid\cI}) &= \hat{B}_{(k,k')\mid\cI}/N\\ 
&=\frac{1}{(\sum_{l\in {\cI}}n_l)^2} \sum_{l\in {\cI}} 
A^3_{lkk'}  + \frac{1}{(\sum_{l\in {\cI}}n_l)^2}\sum_{l, h\in {\cI}}  
A^4_{k'lh}+ \frac{1}{(\sum_{l\in {\cI}}n_l)^2}\sum_{l, h\in {\cI}}  A^4_{klh} 
\\
&\mathrel{\phantom{=}}{} + \frac{2}{(\sum_{l\in {\cI}}n_l)^2}\I(k' \in {\cI})\sum_{l\in {\cI}}  A^5_{k'kl}
+  \frac{2}{(\sum_{l\in {\cI}}n_l)^2} \I(k \in {\cI})\sum_{l\in {\cI}}  A^5_{kk'l} \\
&\mathrel{\phantom{=}}{}
+\frac{1}{\sum_{l\in {\cI}}n_l}\hat{\tau}^2_{(k,k')\mid\cI}
-\frac{2}{(\sum_{l\in {\cI}}n_l)^2} \hat{\tau}_{(k,k')\mid\cI} \{\sum_{l\in {\cI}} n_l (A^1_{lk'}  -
A^1_{lk})\} 
\\
&\mathrel{\phantom{=}}{}
- \I(k' \in {\cI}) \frac{2\hat{\tau}_{(k,k')\mid\cI}}{(\sum_{l\in {\cI}}n_l)^2}  \sum_{l\in {\cI}}n_l A^2_{k'l}  
+ \I(k \in {\cI}) \frac{2\hat{\tau}_{(k,k')\mid\cI}}{(\sum_{l\in {\cI}}n_l)^2}  \sum_{l\in {\cI}}n_l A^2_{kl},
\end{split}
\label{eq:varparametric}
\end{equation}
where 
\begin{equation}
\begin{split}
&
A^3_{lkk'}= \sum_{D_i=l}\{\hat{m}_k(X_i)-\hat{m}_{k'}(X_i)\}^2,  \ A^4_{klh}=  \sum_{D_i=k} \hat{w}_{lk}(X_i) \hat{w}_{hk}(X_i) \{Y_i-\hat{m}_k(X_i)\}^2, \\
& A^5_{kk'l}=  \sum_{D_i=k}  \hat{w}_{lk}(X_i) \{\hat{m}_k(X_i)-\hat{m}_{k'}(X_i)\} \{Y_i-\hat{m}_k(X_i)\}
\end{split}
\label{eq:A15}
\end{equation}
represent three additional types of aggregated data that contribute to the variance estimation.
Various machine learning methods, such as random forests \citep{Wager2018Estimation} and neural networks \citep{Chen1999Improved} can meet the convergence rate requirement of at least $N^{1/4}$ for nuisance function estimation. 
Therefore, our DAC estimator accommodates flexible semiparametric or nonparametric methods for estimating nuisance functions while maintaining the parametric-rate consistency for $\hat\tau_{(k,k')\mid\cI}$. 

\begin{remark}
     DAC estimators using nonparametric methods, which, although asymptotically efficient, can exhibit high variability when the sample size of local single-arm trials is limited; see, for example, the summary of several published applications regarding single-arm studies in Section~3.1 of \citet{phillippo2016nice}. Alternatively, parametric models can be employed. 
\end{remark}

We propose a procedure for implementing the DAC method with nonparametric estimation of nuisance parameters in a distributed network of $K$ distinct single-arm trials. The goal is to estimate ATEs $\tau_{(k,k')\mid\cI}=\E(Y(k') - Y(k){\mid}D\in {\cI})$ across all comparator groups $(k,k')$ for $k, k'\in \{1, \cdots, K\}$ and all target populations ${\cI}\subseteq \{1, \cdots, K\}$ simultaneously. Sites are allowed to interact with a coordinating center by transmitting specific aggregated data.  
Guided by prior knowledge and specific data characteristics, appropriate nonparametric method can be applied to local samples at site $k$ to obtain $\hat{m}_k(X)$, such as random forests and the method of sieves, etc. The method of sieves can be used in the CW approach for estimating weighting functions with general sieve basis functions such as power series, 
splines, wavelets, or artificial neural networks \citep{CHEN2007Large}. 
Each site $j$ is required to provide the aggregated data 
\begin{equation}
\text{AD}_j =\{A^1_{jk}, A^2_{jk}, A^3_{jkk'}, A^4_{jkk'}, A^5_{jkk'}, k, k' \in \{1, \cdots, K\}\}
\label{eq:AD}
\end{equation}
based on Equations (\ref{eq:A12}) and (\ref{eq:A15}) to calculate DAC estimators  $\hat{\tau}_{(k,k')\mid\cI}$  and variance  estimators. 

\begin{algorithm}[!ht]
\caption{DAC with nonparametric methods for estimating working models}
\label{Algorithm}
1. \For{site $j=1$ to $K$}
{
Regress the outcome on covariates at site $j$ to obtain the fitted outcome model function $\hat{m}_j(\cdot)$.\\
Send the summary statistics of covariates $\bar{{g}}_j (X)$ and the fitted outcome model function $\hat{m}_j(\cdot)$ to the coordinating center.\\
}
2. The coordinating center collects and sends $\{\bar{{g}}_1 (X), \hat{m}_1(\cdot),  \bar{{g}}_2 (X), \hat{m}_2(\cdot), \cdots, \bar{{g}}_K (X),  \hat{m}_K(\cdot)\}$  
to all sites.\\
3. \For{site $j=1$ to $K$}
{
Obtain the calibration weights $\hat{w}_{kj}(X_s)/n_k$ via the calibration weighting with $\bar{{g}}_k (X)$ by solving (\ref{eq:weights}) and (\ref{eq:optimize}) 
for $k\neq j$; set $\hat{w}_{jj}(X_s)/n_j=1/n_j$.\\
Compute the aggregated data $\text{AD}_j$ in (\ref{eq:AD}) 
by Equations (\ref{eq:A12}) and (\ref{eq:A15})  
and send them to the coordinating center.\\
}
4. For $k, k' \in \{1, \cdots, K\}$ and ${\cI}\subseteq \{1, \cdots, K\}$, the coordinating center calculates the DAC estimator $\hat{\tau}_{(k,k')\mid\cI}$ of the ATE  by (\ref{eq:DRmultiple}) 
and its variance estimate by  (\ref{eq:varparametric}).
\end{algorithm}

The DAC procedure with nonparametric nuisance parameter estimation is outlined in Figure \ref{fig:workflowDR} and Algorithm \ref{Algorithm}. First, each site $j \in {1, \cdots, K}$ locally regresses the outcome on covariates, and transmits the fitted outcome model function $\hat{m}_j(\cdot)$ and summary statistics of covariates $\bar{{g}}_j(X)$ to the coordinating center. In our algorithm, transmitting a function means sharing the function structure and parameters; for example, a spline method requires sending the order and knots of the spline function, and estimated coefficients of the basis functions.
Second, the coordinating center collects these fitted outcome models and summary statistics, and sends them to each site. Third, each site $j$ uses summary statistics $\bar{{g}}_k(X)$ from other sites $k \neq j$ to perform calibration weighting $K-1$ times to generate $K-1$ sets of weights $\{\hat{w}_{kj}(X_s), D_s=j\}_{k \neq j}$ by solving Equations (\ref{eq:weights}) and (\ref{eq:optimize}). Then aggregated data $\text{AD}_j$ in Equation (\ref{eq:AD})  is computed and sent to the coordinating center.  
Finally, the coordinating center utilizes aggregated data from all sites to compute DAC estimators $\hat{\tau}_{(k,k')\mid\cI}$ of the ATEs and their variance estimators by Equations (\ref{eq:DRmultiple}) and (\ref{eq:varparametric}) for $k, k' \in {1, \cdots, K}$ and ${\cI} \subseteq \{1, \cdots, K\}$.
This algorithm estimates pairwise ATEs across target populations in just two communication rounds, producing identical results to pooled individual-level analysis, as DAC estimators are equivalently expressed in aggregated data form.
\begin{remark}
    Section S5 of the SM presents the distributed outcome regression (DOR) estimation in a distributed network, requiring two communication rounds in four steps. Under linear models, DOR reduces to a single round. In contrast, the DAC approach combines calibration weighting and outcome regression, enhancing robustness without increasing communication costs. 
\end{remark}

\section{Distributed Doubly Robust Inference}
\label{sec:BRDRAW}
Section \ref{sec:DRAWmethod} presents the general DAC method for doubly robust estimation rather than doubly robust inference. In this section, we strengthen the DAC to achieve doubly robust inference under parametric working models.

With parametric working models, the DAC estimator is consistent if either the outcome model or the weighting function is correctly specified. However, the choice of nuisance parameter estimators affects the asymptotic variance of $\hat\tau_{(k,k')\mid\cI}$ under potential model misspecification; see Section S3 of the SM for details.  
Section S4 of the SM provides complex variance estimation for DAC under general parametric nuisance parameter estimation methods with potential misspecification, requiring transmitting more aggregated data as shown in Equation (S.8). 
Fortunately, valid doubly robust inference with parametric working models can be achieved in various contexts \citep{Kim2014doubly,vermeulen2015bias,Tan2020Model,Ning2020robust}, ensuring a generic variance estimation form if at least one working model is correctly specified.   
Let the working model for outcome regression and weighting functions be $\{m_k(X)=m(X; \beta_k): \beta_k \in\R^p\}$ and $\{w_{jk}(X) = w(X; \gamma_{jk}): \gamma_{jk} \in \R^q\}$, respectively. And $\Gamma_{\cI k}= (\gamma_{jk}, j\in\cI)$. Denote nuisance parameter estimators as $\hat{\beta}_k$ and $\hat{\gamma}_{jk}$. And $\hat{\mu}_{k\mid\cI}$ is the DAC estimator in \eqref{eq:dr_mu} using estimated working models $m(X;  \hat{\beta}_k)$ and $w(X; \hat{\gamma}_{jk})$. To achieve doubly robust inference, \citet{vermeulen2015bias} suggest to adapt the nuisance parameter estimation to the EIF by solving the following estimating equations for 
estimates $\hat\beta_k^\br$ and $\hat\Gamma_{\cI k}^\br$,
\begin{equation}\label{eq:br}
\hat\E_N\biggl\{\frac{\partial \varphi_{k\mid\cI}}{\partial\beta_k}(Z; \mu_{k\mid\cI},\hat\beta_{k}^\br,\hat\Gamma_{\cI k}^\br)\biggr\}=0,\quad\hat\E_N\biggl\{\frac{\partial \varphi_{k\mid\cI}}{\partial\gamma_{jk}}(Z; \mu_{k\mid\cI}, \hat\beta_{k}^\br,\hat\Gamma_{\cI k}^\br)\biggr\}=0, j\in\cI,
\end{equation} 
where $\hat\E_Nf(Z)=\sum_{i=1}^Nf(Z_i)/N$ represents empirical mean. The DAC estimator $\hat\mu_{k\mid\cI}^\br$ using $m(X; \hat{\beta}_k^\br)$ and $w(X; \hat{\gamma}_{jk}^\br)$ has a generic asymptotic variance $\E_P\bigl\{\varphi^2_{k\mid\cI}\bigl(Z; \mu^\br_{*k\mid\cI}, \beta_{* k}^\br$, $\Gamma_{*\cI k}^\br\bigr)\bigr\}$, 
where $\mu^\br_{*k\mid\cI}, \beta_{*k}^\br$ and $\Gamma_{*\cI k}^\br$ are probabilistic limits, and the squared first-order asymptotic bias $\text{bias}^2(\beta_{k}$, $\Gamma_{\cI k}; \mu_{k\mid\cI})=[\E\{\varphi_{k\mid\cI}(Z;\mu_{k\mid\cI},\beta_{k},\Gamma_{\cI k})\}]^2$ is minimized in the working model class.

We apply technique \eqref{eq:br} to the DAC under log-linear weighting functions and linear outcome models, i.e., $m_\ell(X; \beta_{\ell})=\beta_{\ell}^Tg(X)$ and $w_{j\ell}(X; \gamma_{j\ell})=\exp\bigl(\gamma_{j\ell}^T a(X)\bigr)$ for $\ell=k,k'$. Components in both sets of basis functions $g(X)=\{g_1(X),\dots,g_p(X)\}^T$ and $a(X)=\{a_1(X),\dots,a_q(X)\}^T$ are linearly independent. Then, nuisance parameters are estimated by solving 
\begin{align}
\frac{1}{N}\sum_{i=1}^N I(D_i=j) g(X_i) - \frac{1}{N}\sum_{i=1}^N I(D_i=\ell)\exp\bigl(\gamma_{j\ell}^T a(X_i)\bigr)g(X_i) &=0, \label{eq:loglinearBR}\\
\frac{1}{N}\sum_{i=1}^N\biggl\{I(D_i=\ell)\sum_{j\in\cI}\exp\bigl(\gamma_{j\ell}^Ta(X_i)\bigr)\bigl\{Y_i-\beta_\ell^Tg(X_i)\bigr\}a(X_i)\biggr\}&=0. \label{eq:linearBR}
\end{align}
Identification of nuisance parameters requires $p=q$ and full-rank conditions on the matrices  $\E\{I(D=\ell)\sum_{j\in\cI} \exp\bigl(\gamma_{j\ell}^T a(X)\bigr) a(X) g^T(X) \}$ for $\cI \subseteq \{1, \cdots, K\}$. 
Equation \eqref{eq:loglinearBR} aligns with the CW method to balance $g(X)$. The solution to \eqref{eq:loglinearBR} and \eqref{eq:linearBR} by generalized method of moments (GMM) is denoted as $\hat{\beta}^{\br}_{\ell \mid \cI}$ and $\hat\Gamma_{\cI\ell}^\br=(\hat\gamma_{j\ell}^\br,j\in\cI)$.
Then, the DAC estimator $\hat\mu_{\ell\mid\cI}^\br$  is the estimator in Equation (\ref{eq:dr_mu}) 
with $\hat w_{j\ell}(X)=\exp\{(\hat\gamma_{j\ell}^\br)^Ta(X)\}$ and $\hat m_\ell(X)=(\hat{\beta}^{\br}_{\ell \mid \cI})^Tg(X)$, and 
the variance estimate for $\hat{\tau}_{(k,k')\mid\cI}^\br=\hat\mu_{k'\mid\cI}^\br-\hat\mu_{k\mid\cI}^\br$ is $\hat\E_N\bigl\{\phi^2_{(k,k')\mid\cI}\bigl(Z;\hat{\tau}_{(k,k')\mid\cI}^\br, \hat{\beta}^{\br}_{\ell \mid \cI},\hat\Gamma_{\cI \ell}^\br, \ell=k,k'\bigr)\bigr\}/N$.

\begin{remark}
    Notably, if $g(X)=a(X)$, Equation \eqref{eq:loglinearBR} aligns with the entropy balancing method to balance $g(X)$, suggesting solving Equation (\ref{eq:optimize}) to obtain the estimate $\hat{\gamma}^\br_{j\ell}$.
    Equation \eqref{eq:linearBR} indicates performing weighted least squares with weights $\sum_{j\in\cI}\exp\{(\hat\gamma_{j\ell}^\br)^T a(X_i)\}$ at site $l$ to obtain $\hat\beta_{\ell \mid \cI}^\br$. And $\hat{\tau}^{\br}_{(k,k')\mid\cI}$ simplifies to an outcome regression estimator with $\hat m_\ell(X)=\{\hat\beta_{\ell \mid \cI}^\br\}^Tg(X)$, as components $\{A^2_{lj}, l \in \{k, k'\}, j\in \cI\}$ of the bias correction term in \eqref{eq:A12} are weighted averages of residuals from weighted least squares which equal zero. From Equation \eqref{eq:DACCWdif}, the DAC estimator $\hat{\tau}^{\br}_{(k,k')\mid\cI}$ is also equivalent to the DCW estimator $\hat{\tau}_{(k,k')\mid\cI}^{\text{DCW}}$ with $g(X)$ to be calibrated, as balancing $g(X)$ automatically balances $\hat{m}_k(X)$ under linear outcome models. Thus, in this case, Equation \eqref{eq:linearBR} is actually unnecessary for the point estimation of $\tau_{(k,k')\mid\cI}$, as the DAC estimator with $\hat w_{j\ell}(X)=\exp\{(\hat\gamma_{j\ell}^\br)^Tg(X)\}$ remains the DCW estimator regardless of how $\beta_\ell$ is estimated. This finding not only confirms doubly robust consistency \citep{Zhao2016Entropy} but also reveals the doubly robust inference property of the entropy balancing estimator with respect to the linear outcome regression model and log-linear weighting function with same basis functions.
\end{remark}

\begin{figure}[!bht]
\centering
\includegraphics[width=0.95\textwidth]{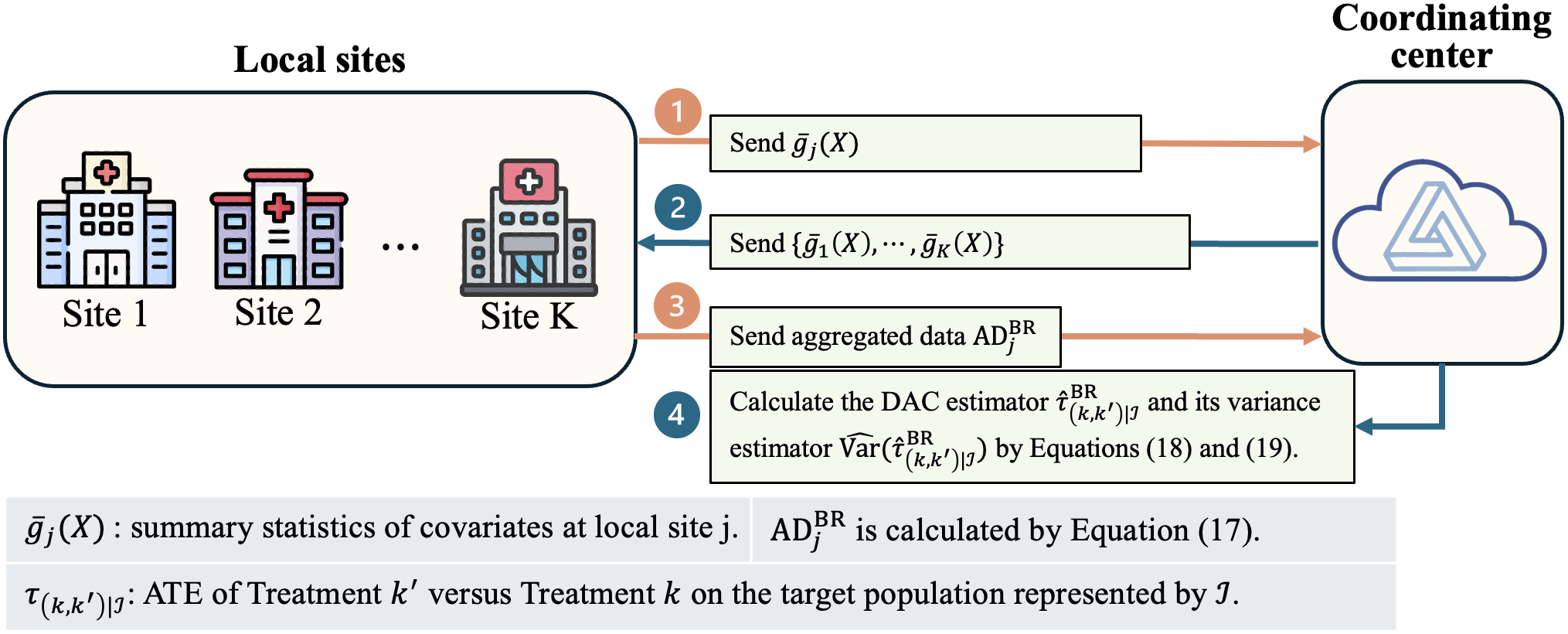}
\caption{Workflow of the distributed augmented calibration weighting (DAC) estimation under a log-linear weighting function and linear outcome regression model to achieve doubly robust inference in a distributed network of multiple single-arm trials.
\label{fig:workflowBR}
}
\end{figure}

Let the aggregated data from site $j$ be
\begin{equation}
\text{AD}^{\br}_j =\{\hat{\beta}^{\br}_{j \mid \cI}, O^1_{j}, O^2_{jk \mid \cI}, O^3_{j}, O^4_{jkk' \mid \cI}, O^5_{jk \mid \cI}, k, k' \in \{1, \cdots, K\}, \cI \subseteq \{1, \cdots, K\}\},
\label{eq:ADBR}
\end{equation}
where  
$O_j^1 = \sum_{i:D_i=j}g(X_i)$, $O^2_{jk\mid \cI}= \sum_{i:D_i=j} \exp\{\hat\gamma_{kj}^T a(X_i)\} \{Y_i-(\hat{\beta}^{\br}_{j \mid \cI})^Tg(X_i)\}$, 
    $O^3_{j}=\sum_{i:D_i=j}$ $g(X_i)g^T(X_i)$, $O^4_{jlh \mid \cI}  =  \sum_{i:D_i=j} \exp\{(\hat\gamma_{lj} + \hat\gamma_{hj})^T a(X_i)\}\{Y_i- (\hat{\beta}^{\br}_{j \mid \cI})^T g(X_i)\}^2$ and $O^5_{jk\mid \cI}=  \sum_{D_i=j}  \exp\{\hat\gamma_{kj}^T a(X_i)\} \{Y_i-(\hat{\beta}^{\br}_{j \mid \cI})^Tg(X_i)\} g(X_i)$.
The DAC estimator in \eqref{eq:DRmultiple} under log-linear  weighting functions and linear outcome models, along with its variance estimator, simplifies to 
\begin{align}
\hat{\tau}^{\br}_{(k,k')\mid\cI} 
&= \frac{1}{\sum_{j\in \mathcal{I}}n_j} \{(\hat{\beta}^{\br}_{k' \mid\cI}  -
\hat{\beta}^{\br}_{k \mid\cI})^T   (\sum_{j \in \mathcal{I}}   O_j^1)+ \sum_{j \in \mathcal{I}} O^2_{k'j \mid \cI} - \sum_{j \in \mathcal{I}} O^2_{kj \mid \cI}\}, \label{eq:BRDRAW}\\ 
\widehat{\V}(\hat{\tau}^{\br}_{(k,k')\mid\cI})
&=\frac{2}{(\sum_{l\in {\cI}}n_l)^2} \hat{\tau}^{\br}_{(k,k')\mid\cI} (\hat{\beta}^{\br}_{k \mid\cI}  -
\hat{\beta}^{\br}_{k' \mid\cI})^T(\sum_{l\in {\cI}} O^1_{l}) + \frac{1}{\sum_{l\in {\cI}}n_l}\{\hat{\tau}^{\br}_{(k,k')\mid\cI}\}^2  \label{eq:BRDRAWvar}
\\
&\mathrel{\phantom{=}}{} +
 \frac{2\hat{\tau}^{\br}_{(k,k')\mid\cI}}{(\sum_{l\in {\cI}}n_l)^2}  \{\I(k \in {\cI}) \sum_{l\in {\cI}}O^2_{kl \mid \cI} - \I(k' \in {\cI}) \sum_{l\in {\cI}}O^2_{k'l \mid \cI}\}, \nn \\
&\mathrel{\phantom{=}}{} +\frac{1}{(\sum_{l\in {\cI}}n_l)^2} (\hat{\beta}^{\br}_{k' \mid\cI}  -
\hat{\beta}^{\br}_{k \mid\cI})^T (\sum_{l\in {\cI}} 
O^3_{l}) (\hat{\beta}^{\br}_{k' \mid\cI}  -
\hat{\beta}^{\br}_{k \mid\cI})  + \frac{\sum_{l, h\in {\cI}}  
(O^4_{k'lh \mid\cI}+ O^4_{klh \mid\cI})}{(\sum_{l\in {\cI}}n_l)^2} \nn \\
&\mathrel{\phantom{=}}{} + \frac{2}{(\sum_{l\in {\cI}}n_l)^2} (\hat{\beta}^{\br}_{k' \mid\cI}  -
\hat{\beta}^{\br}_{k \mid\cI})^T \{\I(k' \in {\cI})\sum_{l\in {\cI}} O^5_{k'l \mid \cI} 
-\I(k \in {\cI})\sum_{l\in {\cI}}  O^5_{kl \mid \cI}\}. \nn
\end{align}
The variance estimation in \eqref{eq:BRDRAWvar} holds even if one working model is misspecified.
Figure \ref{fig:workflowBR} and Algorithm \ref{algorithmBR} outline a four-step DAC estimation procedure with two communication rounds for this case. 
For variable selection of $a(X)$ and $g(X)$, we can consider variables that exhibit distinct distributions across sites and are important for local outcome models.

\begin{algorithm}[!ht]
\caption{DAC under a log-linear weighting function and a linear outcome regression model to achieve doubly robust inference}
\label{algorithmBR}
1. \For{site $j=1$ to $K$}
{
Send the summary statistics of covariates $\bar{{g}}_j (X)$ to the coordinating center.\\
}
2. The coordinating center collects and sends $\{\bar{{g}}_1 (X), \bar{{g}}_2 (X), \cdots, \bar{{g}}_K (X)\}$  
to all sites.\\
3. \For{site $\ell=1$ to $K$}
{
Solve \eqref{eq:loglinearBR} and \eqref{eq:linearBR} by GMM to obtain $\hat{\gamma}^\br_{j\ell}$ and $\hat{\beta}^{\br}_{\ell \mid \cI}$ for $j\in \{1, \cdots, K\}$ and ${\cI}\subseteq \{1, \cdots, K\}$, using local data at site $\ell$ and $\{\bar{{g}}_1 (X), \bar{{g}}_2 (X), \cdots, \bar{{g}}_K (X)\}$.\\
Compute the aggregated data $\text{AD}^{\br}_\ell$ in (\ref{eq:ADBR}) and send them to the coordinating center.\\
}
4. For $k, k' \in \{1, \cdots, K\}$ and ${\cI}\subseteq \{1, \cdots, K\}$, the coordinating center calculates the DAC estimator $\hat{\tau}^{\br}_{(k,k')\mid\cI}$ of ATE  by \eqref{eq:BRDRAW} and its variance estimate by \eqref{eq:BRDRAWvar}.
\end{algorithm}

Generally, under generalized linear working models, $m_\ell(X)$ and $w_{j\ell}(X)$ for $\ell=k,k'$ are modeled as $\bigl\{h_m\bigl(\beta^T_\ell g(X)\bigr):\beta \in[-M,M]^p\bigr\}$ and $\bigl\{h_w\bigl(\gamma ^T_{j\ell} a(X)\bigr): \gamma_{j\ell} \in[-M,M]^q\bigr\}$, respectively, where $h_m(\cdot)$ and $h_w(\cdot)$ are link functions, and $M$ is a large positive constant to ensure compactness of the parameter space. 
This setup supports various outcome types, including continuous and categorical data, by appropriate choices of $h_m(\cdot)$. To ensure valid weights, $h_w(\cdot)$ is constrained to be positive. 
The technique in \eqref{eq:br} motivates the following estimating equations:
\begin{align}
\frac{1}{N}\sum_{i=1}^N h_m'\bigl(\beta_\ell^Tg(X_i)\bigr)g(X_i)\bigl\{I(D_i=j)-I(D_i=\ell)h_w\bigl(\gamma_{j\ell}^T a(X_i)\bigr)\bigr\}&=0\label{eq:ee_weight}\\
\frac{1}{N}\sum_{i=1}^N\biggl\{I(D_i=\ell)\sum_{j\in\cI}h_w'\bigl(\gamma_{j\ell}^T a(X_i)\bigr)\bigl\{Y_i-h_m\bigl(\beta_\ell^Tg(X_i)\bigr)\bigr\} a(X_i)\biggr\}&=0,\label{eq:ee_outcome}
\end{align}
where $j\in\cI$ and $h_w'(\cdot)$ denotes the first-order derivative of $h_w(\cdot)$. And some conditions (Assumption 7 in the SM) are required for the identification of nuisance parameters in above estimating equations.
Equation \eqref{eq:ee_weight} corresponds to calibration weighting with variables $h_m'\bigl(\beta_\ell^Tg(X)\bigr)g(X)$ to be calibrated. 
The solution to \eqref{eq:ee_weight} and \eqref{eq:ee_outcome} is $(\hat\beta_\ell^\br,\hat\gamma_{j\ell}^\br,j\in\cI)$.  
Then the bias-reduced DAC estimator $\hat\mu_{\ell\mid\cI}^\br$  is  
the estimator in Equation (\ref{eq:dr_mu}) 
with estimated working models $\hat{w}_{j\ell}(X) = h_w\bigl(\{\hat\gamma_{j\ell}^\br\}^T a(X)\bigr)$ and $\hat{m}_\ell(X) = h_m\bigl(\{\hat\beta_\ell^\br\}^Tg(X)\bigr)$.
As stated in Theorem \ref{thm:dr_para}, the treatment effect estimator $\hat{\tau}_{(k,k')\mid\cI}^\br = \hat\mu_{k'\mid\cI}^\br-\hat\mu_{k\mid\cI}^\br$ is asymptotically normal with a simple and generic form of asymptotic variance, even when one of the working models is misspecified. Moreover, it minimizes the squared first-order asymptotic bias when both models are misspecified.  
For complex models, solving Equations \eqref{eq:ee_weight} and \eqref{eq:ee_outcome} may require iterative methods, and developing a communication-efficient computation scheme is left for future work. 

\begin{remark}
    The DAC estimator introduced in Section \ref{sec:DRAWmethod} is a general estimator accommodating both parametric and nonparametric estimation of the two nuisance working models. The bias-reduced DAC estimator is a specialized version under parametric nuisance models, achieving doubly robust inference with minimal squared first-order asymptotic bias. Its key technique is to estimate nuisance parameters by enforcing that the derivative of the efficient influence function equals zero (Equations \eqref{eq:loglinearBR} and \eqref{eq:linearBR}), which makes outcome model estimation depend on calibration weights and thus requires weights to be estimated first. A direct implementation of Algorithm \ref{Algorithm} using the aggregated data in Equation \eqref{eq:AD} would require an extra communication round. To avoid this, we rewrite estimators with an alternative aggregated data structure in Equation \eqref{eq:ADBR}, and perform bias-reduced DAC using Algorithm \ref{algorithmBR} with two communication rounds. In summary, Algorithm \ref{Algorithm} applies to nonparametric nuisance estimation, whereas Algorithm \ref{algorithmBR} enables bias-reduced inference under parametric models with  minimal communication cost.
\end{remark}

\section{Asymptotic Properties}
\label{sec:theory}
This section formally illustrates the EIFs for target estimands and asymptotic properties of DAC and bias-reduced DAC estimators. The following theorem presents the EIFs of $\mu_{k\mid\cI}$  and $\tau_{(k,k')\mid\cI}$. 
\begin{theorem}\label{pro:2}
Under Assumptions~\ref{as:1}–\ref{as:3}, the efficient influence function for $\mu_{k\mid\cI}$ is 
\begin{equation*} 
\varphi_{k\mid\cI}\bigl(Z;P\bigr)=\frac{1}{\pr(D\in\cI)}\biggl[I(D\in\cI)\bigl\{m_k(X)-\mu_{k\mid\cI}\bigr\}+I(D=k)\bigl\{Y-m_k(X)\bigr\}\sum_{j\in\cI}w_{jk}(X)\biggr], 
\end{equation*}
where $P$ in $\varphi_{k\mid\cI}\bigl(Z;P\bigr)$ indicates that all the parameters $\mu_{k\mid\cI}$, $m_k(X)$, and $w_{jk}(X)$ are defined under the distribution $P$.
The efficient influence function for $\tau_{(k,k')\mid\cI}$ is
\begin{equation*}
\phi_{(k,k')\mid\cI}\bigl(Z;P\bigr)=\varphi_{k'\mid\cI}\bigl(Z;P\bigr)-\varphi_{k\mid\cI}\bigl(Z;P\bigr).
\end{equation*}
\end{theorem}

\begin{remark}
    Extensions for EIFs in cases of repeated treatments across different sites or multiple treatments assigned within a local site are detailed in Section S6 of the SM. 
\end{remark}
To formally state doubly robust consistency of the proposed distributed estimator, let $\cF$ and $\cG$ be the working model classes for estimating outcome regression and weighting function, respectively. The squared $L_2(P)$-norm is denoted as $\|f\|_P^2=\E_P\{f(Z)\}^2$. Let $w_{\cI\ell}=(w_{j\ell},j\in\cI)^T$.

\begin{theorem}\label{prop:dr_consistency}
Under Assumptions~\ref{as:1}--\ref{as:3} and some additional technical conditions (Assumptions 4 and 5 in Section S1.3 of the SM), the DAC estimator $\hat\tau_{(k,k')\mid\cI}$ in \eqref{eq:DRmultiple} based on $\hat{m}_\ell$ and $\hat{w}_{j\ell}$ has doubly robust consistency with respect to $(\cF,\cG)$ in the sense that it is consistent if either $m_\ell(X)=m^*_\ell(X)$ or $w_{\cI\ell}(X)=w^*_{\cI\ell}(X)$ holds for $\ell=k$ and $k'$, where $m^*_\ell(X)$ and $w^*_{jl}(X)$ are probabilistic limits of $\hat{m}_\ell(X)$ and $\hat{w}_{j\ell}(X)$, i.e., $\|\hat{m}_\ell-m^*_\ell\|_P {\to_P} 0$ and $\|\hat{w}_{j\ell}-w^*_{j\ell}\|_P {\to_P} 0$.
\end{theorem} 
 Under some assumptions on convergence rates for working models, the following theorem establishes the basis for inference on the DAC estimator $\hat\tau_{(k,k')\mid\cI}$ in \eqref{eq:DRmultiple}.   
\begin{theorem}\label{prop:npara}
Under Assumptions~\ref{as:1}--\ref{as:3} and additional conditions (Assumptions 4--6 in the SM), 
\[
\sqrt{N}\bigl\{\hat\tau_{(k,k')\mid\cI}-\tau_{(k,k')\mid\cI}\bigr\}\to_dN(0,B_{(k,k')\mid\cI})
\]
if,
for $\ell=k,k'$, $\|\hat{m}_\ell-m_\ell\|_P=o_{\pr}(1)$, $\|\hat{w}_{\cI\ell}-w_{\cI\ell}\|_P=o_{\pr}(1)$, and $\|\hat{m}_\ell-m_\ell\|_P\times\|\hat{w}_{\cI\ell}-w_{\cI\ell}\|_P=o_{\pr}\bigl(N^{-1/2}\bigr)$. Moreover, $B_{(k,k')\mid\cI}$ can be consistently estimated by the sample second moment of the efficient influence function; that is, $\hat{B}_{(k,k')\mid\cI}{\to_P} B_{(k,k')\mid\cI}$ with
$\hat{B}_{(k,k')\mid\cI}=\frac{1}{N}\sum_{i=1}^N\bigl\{\phi_{(k,k')\mid\cI}(Z_i;\hat{P})\bigr\}^2$,
where $\hat{P}$ represents that nuisance parameters in $\phi_{(k,k')\mid\cI}$ are plugged in by their respective estimators.
\end{theorem}

We demonstrate that the bias-reduced DAC estimator $\hat{\tau}_{(k,k')\mid\cI}^\br$ under generalized linear models exhibits the doubly robust inference and bias reduction properties in the following theorem.
\begin{theorem}\label{thm:dr_para}
Under Assumptions~\ref{as:1}--\ref{as:3} and some additional technical conditions (Assumptions 4, 6 and 7 in the SM), the following two statements hold: The solution to \eqref{eq:ee_weight} and \eqref{eq:ee_outcome} is $\hat\beta_\ell^\br$ and $\hat\Gamma_{\cI\ell}^\br=(\hat\gamma_{j\ell}^\br,j\in\cI)$, and $\beta_{* \ell}^\br$ and $\Gamma_{*\cI \ell}^\br$ are their probabilistic limits for $\ell \in \{k,k'\}$.
\begin{enumerate}[label=(\arabic*)]
\item If
for $\ell=k,k'$, either $m_\ell(X)=h_m\bigl(\beta_{\ell}^Tg(X)\bigr)$ or $w_{j\ell}(X)=h_w\bigl(\gamma_{j\ell}^T a(X)\bigr)$ holds for some $\beta_{\ell}\in[-M,M]^p, \gamma_{j\ell}\in[-M,M]^q$, then 
\[
\sqrt{N}\bigl\{\hat{\tau}_{(k,k')\mid\cI}^\br-\tau_{(k,k')\mid\cI}\bigr\}\to_dN(0, \E_P\bigl\{\phi^2_{(k,k')\mid\cI}\bigl(Z; \tau_{(k,k')\mid\cI}, \beta_{* \ell}^\br, \Gamma_{*\cI \ell}^\br, \ell=k,k'\bigr)\bigr\}),
\]
and the asymptotic variance can be consistently estimated by
$\frac{1}{N}\sum_{i=1}^N \bigl\{\phi_{(k,k')\mid\cI}\bigl(Z_i;\hat{\tau}_{(k,k')\mid\cI}^\br$, $\hat{\beta}^{\br}_{\ell},\hat\Gamma_{\cI \ell}^\br, \ell=k,k'\bigr)\bigr\}^2$.

\item The squared first-order asymptotic bias of $\hat{\tau}_{(k,k')\mid\cI}$ under working models $\bigl\{h_m\bigl(\beta^T_\ell g(X)\bigr):\beta_{\ell} \in[-M,M]^p\bigr\}$ and $\bigl\{h_w\bigl(\gamma ^T_{j\ell} a(X)\bigr): \gamma_{j\ell} \in[-M,M]^q\bigr\}$ is 
\[\text{bias}^2(\beta_{\ell},\Gamma_{\cI \ell}, \ell=k,k'; \tau_{(k,k')\mid\cI})=[\E_P\{\phi_{(k,k')\mid\cI}(Z;\tau_{(k,k')\mid\cI},\beta_{\ell},\Gamma_{\cI \ell}, \ell=k,k')\}]^2.\]
We have $\text{bias}^2(\beta_{*\ell}^\br,\Gamma_{*\cI \ell}^\br, \ell=k,k'; \tau_{(k,k')\mid\cI}) \leq \text{bias}^2(\beta_{\ell},\Gamma_{\cI \ell}, \ell=k,k'; \tau_{(k,k')\mid\cI})$ for all $\beta_{\ell}$ and $\Gamma_{\cI \ell}$ within the parameter space $(-M,M)^{2q\times|\cI|+2p}$.
\end{enumerate}
\end{theorem}

\section{Simulation} 
\label{sec:simulation} 

We conduct simulations to evaluate the finite sample performance of DAC estimators with linear outcome models and entropy balancing (Figure \ref{fig:workflowBR}), and compare them to DOR estimators which use ordinary least squares to estimate coefficients of linear models (Figure S3 in the SM).
The number of sites in the distributed network is set to $K = 4$, with a total sample size of  $N \in \{800, 2400, 4000\}$ subjects across all sites. Baseline covariates $X = (X_1, X_2, X_3)^T$ are generated by a multivariate normal distribution with mean $(0.6, 0.6, 0.6)$, variances $(0.64, 1, 1.44)$ and correlations $\rho_{X_1 X_2} = \rho_{X_2 X_3} = 0.001$, $\rho_{X_1 X_3} = 0$. The site index $D$ is generated by the following two strategies using the multinoulli distribution with the parameters of probabilities being
\begin{align*}
&P(D=k \mid  X)= \exp\{\zeta_{k1} X_1+\zeta_{k2} X_2 +\zeta_{k3} X_3\}/ \sum_{i=1}^K \exp\{\zeta_{i1} X_1+\zeta_{i2} X_2 +\zeta_{i3} X_3\} \text{ and }\\
&P(D=k \mid  X)= \exp\{\nu_{k0} + \nu_{k1} X_1^2+\nu_{k2} X_2^2 +\nu_{k3} X_3^2 \}/ \sum_{i=1}^K \exp\{\nu_{i0} + \nu_{i1} X_1^2 +\nu_{i2} X_2^2 +\nu_{i3} X_3^2 \},
\end{align*}
respectively, where $\zeta_{11}=0.2, \zeta_{12}=-0.2, \zeta_{13} = -0.1, 
\zeta_{21}=0.36, \zeta_{22}= -0.28, \zeta_{23} = -0.16, 
\zeta_{31}= 0.5, \zeta_{32}= -0.4, \zeta_{33} = -0.2,
\zeta_{41}= 0.65, \zeta_{42}= -0.5, \zeta_{43} = -0.25,   
\nu_{10} = -0.4, \nu_{11} = 1.6, \nu_{12} = 1, \nu_{13} = 0.25,  
\nu_{20} = -0.3, \nu_{21} = 1.4, \nu_{22} = 0.7, \nu_{23} = 0.5, 
\nu_{30} = -0.5, \nu_{31} = 1.55, \nu_{32} = 1.1, \nu_{33} = 0.3, 
\nu_{40} = -0.1, \nu_{41} = 1.5, \nu_{42} = 0.6, \nu_{43} = 0.4$.
Under both strategies, each site accounts for approximately 25\% of the overall population. Similarly,  two strategies are used to generate the observed outcomes $Y$ with  normal distribution,
\begin{align*}
&Y \mid (X, D=k) \sim N(\theta_{k0} + \theta_{k1} X_1+\theta_{k2} X_2 +\theta_{k3} X_3, 0.04 |X_2|^{0.4} ) \text{ and }\\
&Y \mid  (X, D=k) \sim N(\psi_{k0} + \psi_{k1} X^2_1+\psi_{k2} X^2_2 +\psi_{k3} X^2_3, 0.04 |X_2|^{0.4} ),
\end{align*}
respectively, where $\theta_{10}=1, \theta_{11} = 0.3, \theta_{12} = 0.2, \theta_{13} = 0.2, 
\theta_{20}= 4, \theta_{21} = 0.5, \theta_{22} = 0.5, \theta_{23} = 0.6, 
\theta_{30}= 7, \theta_{31} = 0.7, \theta_{32} = 0.8, \theta_{33} = 1, 
\theta_{40}= 10, \theta_{41} = 0.9, \theta_{42} = 1.1, \theta_{43} = 1.5, 
\psi_{10} = -0.5, \psi_{11} = -1, \psi_{12} = -1, \psi_{13} = -0.5,
\psi_{20} = 0.2, \psi_{21} =  0.5, \psi_{22} = 1, \psi_{23} = 0.5,
\psi_{30} = 1.5, \psi_{31} =  1, \psi_{32} = 2, \psi_{33} = 1,
\psi_{40} = 3, \psi_{41} =  1.5, \psi_{42} = 2.5, \psi_{43} = 2$.
We evaluate two estimation methods under four data-generating scenarios outlined in Table \ref{tb:estimators} to estimate $\tau_{(k,k')\mid\cI}$, simulating 600 replicates for each scenario. 

\begin{table}[!ht]
    \caption{Descriptions of (a)  four   data-generating scenarios in simulations, and  (b)  two estimation methods, their types, the random vector $g(X)$ to be calibrated in the calibration weighting and  the working model $m_k(X)$ for outcome regression. A ``---" indicates the working model is not involved in the estimation method. \label{tb:estimators}}
    \centering
    {(a)  Data-generating scenarios}\\
     \scalebox{0.77}{
    \begin{tabular}{cc}
    \hline
        Scenario & Details\\
         \hline 
       (i)  & generating $D$ and $Y$ both by the first strategy, 
respectively\\
       (ii)  & generating $D$ by the first strategy but $Y$ by the second\\ 
       (iii) & generating $D$ by the second strategy while $Y$ by the first\\
        (iv) & generating $D$ and $Y$ both by the second strategy, 
respectively\\
         \hline
    \end{tabular}} 
    \\
		\medskip
		\centering
		{(b)  Estimation methods}\\
        \scalebox{0.77}{
    \begin{tabular}{cccc}
    \hline
        Method & Estimator type   &  Calibrated random vector ${g}(X)$   & Outcome  model $m_k(X)$\\
         \hline 
        DOR  & outcome regression &  --- &  
       linear model \\
        DAC & augmented calibration weighting  & $(X_1, X_2, X_3)^T$ &  linear model \\
         \hline
    \end{tabular}} 
    
\end{table}

\begin{figure}[!ht]
\centering
\includegraphics[width=0.7\textwidth]{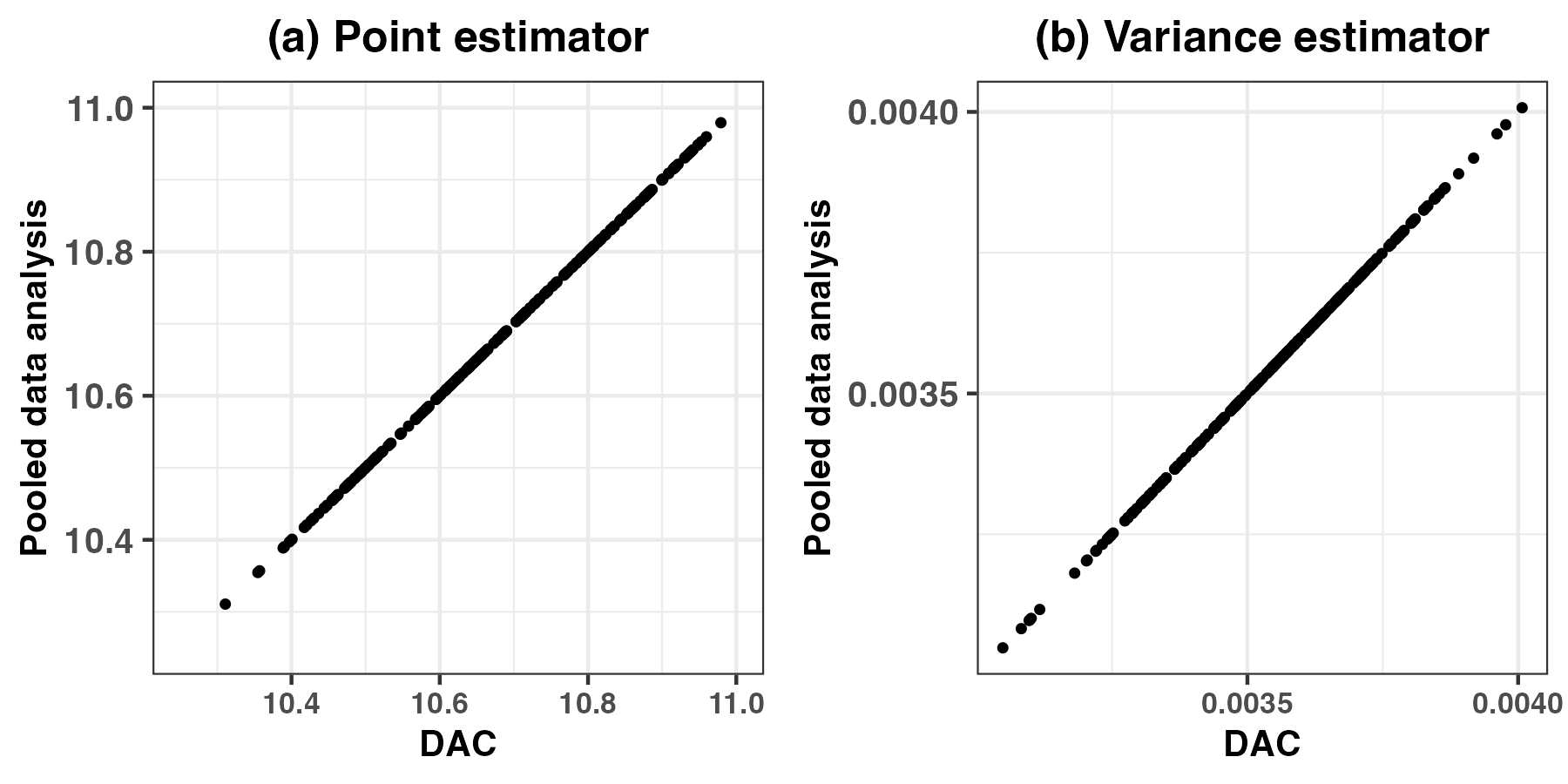}\\		
\caption{Scatterplots of (a) point and (b)  variance estimates from the  augmented calibration weighting estimation obtained in distributed setting (DAC) and   in  pooled data analysis. \label{figure:lossless}
}
\end{figure}

\begin{table}[!ht]
\centering
\caption{Simulation results of comparing  Treatment 4 versus Treatment 1 on the overall population in a distributed network with four single-arm trials under three different values of the total sample size $N$ based on 600 Monte Carlo replications. \label{tb:simulation}
}
\scalebox{0.75}{
\begin{tabular}{c|c|cccc|cccc|cccc}
\hline
Scenario &  & \multicolumn{4}{c|}{$N=800$}  & \multicolumn{4}{c|}{$N=2400$} & \multicolumn{4}{c}{$N=4000$}\\
True value &  Method & Bias$\times10^3$ & SD & ESE & CP & Bias$\times10^3$ & SD & ESE & CP & Bias$\times10^3$ & SD & ESE & CP \\ 
\hline
(i) 10.68 
 & DOR & -0.57 & 0.06 & 0.07 & 97.17 & -0.83 & 0.04 & 0.04 & 94.33 & -0.29 & 0.03 & 0.03 & 96.33 \\  
 & DAC & -0.59 & 0.06 & 0.07 & 97.17 & -0.83 & 0.04 & 0.04 & 94.67 & -0.28 & 0.03 & 0.03 & 96.17 \\ 
 \hline
(ii) 15.26  
 & DOR & -273.60 & 0.47 & 0.48 & 89.83 & -258.13 & 0.29 & 0.28 & 81.83 & -237.96 & 0.22 & 0.22 & 78.33 \\  
 & DAC & -9.33 & 0.47 & 0.48 & 95.50 & -10.86 & 0.29 & 0.28 & 94.00 & 4.45 & 0.22 & 0.22 & 94.83 \\ 
 \hline
(iii) 10.68  
 & DOR & 0.17 & 0.07 & 0.07 & 95.33 & 0.61 & 0.04 & 0.04 & 95.17 & 0.35 & 0.03 & 0.03 & 94.17 \\  
 & DAC & 0.17 & 0.07 & 0.07 & 95.50 & 0.60 & 0.04 & 0.04 & 94.50 & 0.34 & 0.03 & 0.03 & 94.17 \\ 
 \hline
(iv) 15.27 
 & DOR & -1112.65 & 0.46 & 0.44 & 31.17 & -1059.41 & 0.26 & 0.25 & 1.50 & -1048.71 & 0.20 & 0.20 & 0.17 \\ 
 & DAC & -803.27 & 0.46 & 0.44 & 54.33 & -770.35 & 0.27 & 0.26 & 17.17 & -760.74 & 0.20 & 0.20 & 4.17 \\ 
\hline
\end{tabular}}
{\raggedright 
Note: Scenarios and Methods  are described in Table \ref{tb:estimators}. Bias (magnified 1000 times for presentation) and SD are the Monte Carlo bias and standard deviation across the 600 simulations of the points estimates. ESE and CP(\%) are the averages of estimated asymptotic standard errors and coverage proportions of 95\% confidence intervals based on the plug-in method, respectively. 
 \par}
\end{table}

Figure \ref{figure:lossless} shows that augmented calibration weighting estimates calculated by aggregated data in the distributed setting are identical those from pooled data analysis, confirming the lossless property of the DAC method. 
Table \ref{tb:simulation} summarizes the performance of DAC estimators and DOR estimators in estimating ATEs for treatment 4 versus treatment 1 on the overall population $\{1, 2, 3, 4\}$. Results for  target populations $\{3\}, \{1, 3\}, \{1, 3, 4\}$ are detailed in Tables S1-S3 in the SM.
For DAC estimators,  the averages of estimated asymptotic standard errors (ESEs) are close to the Monte Carlo standard deviations (SDs), validating the asymptotic variance estimation and the doubly robust inference property. 
DAC estimators with at least one correctly specified working model under Scenarios (i)-(iii) exhibit very small bias and comparable Monte Carlo standard deviations relative to DOR estimators, even for small sample sizes. In contrast, DOR estimators in Scenarios (ii) and (iv) display large bias when their working models are misspecified. DAC reduces DOR's bias by 25.3\%–31.3\% when neither working model is correctly specified. Coverage proportions based on DAC estimators approximate the nominal level of 0.95 when at least one working model is correctly specified in Scenarios (i)-(iii), while OR estimators achieve nominal coverage only under correctly specified outcome models.

\section{Real Data Analysis}
\label{sec:empirical} 
We conducted our empirical analysis using data from the Magnetic Resonance Imaging (MRI) sub-study of the Memory in Diabetes (MIND) substudy within the Action to Control Cardiovascular Risk in Diabetes (ACCORD) trial. ACCORD was a randomized trial of 10,251 middle-aged and older participants with type 2 diabetes mellitus at high cardiovascular risk due to prevalent cardiovascular disease or additional risk factors. Participants in the main ACCORD trial were randomized to either a systolic blood pressure goal of less than 120 mm Hg (intensive BP therapy)  versus less than 140 mm Hg (standard BP therapy), or to a fibrate versus placebo for patients with low-density lipoprotein cholesterol levels less than 100 mg/dL.  
632 participants from ACCORD trial who agreed to MRI procedure were enrolled in MRI substudy to examine whether intensive hypertension therapy and combination therapy with a statin plus fibrate could reduce the risk of total brain volume (TBV) decline in patients with type 2 diabetes mellitus. A flow chart in Figure S5 of the SM shows the cohort participation; see \cite{Williamson2014cognitive} for details on the study design.  
Of these participants, 614  completed the baseline TBV measurement within 45 days after randomization, 
and 503 completed a 40-month follow-up TBV measurement.
This final subset comprised 314 participants from the blood pressure (BP) trial (161 standard BP and 153 intensive BP) and 189 from the lipid trial (100 placebo and 89 fibrate). Due to participant attrition during follow-up, there were some imbalances in baseline characteristics of these 503 participants, as shown in Figure S6 of the SM. 
\cite{Williamson2014cognitive} analyzed these patient-level data data using analysis of covariance with the adjustment on baseline characteristics, equivalent to outcome regression estimation with linear models,  reporting a 4.4 cm$^3$ [95\% CI, 1.1 to 7.8 cm$^3$; P-value = 0.01] greater TBV decline in the intensive BP group compared to the standard BP group at 40 months, while fibrate therapy had no effect on TBV changes compared to placebo.

Here we investigate effects of the four interventions (standard BP, intensive BP, lipid placebo and lipid fibrate) using DAC method (Figure \ref{fig:workflowDR}),   
treating each treatment group as a single-arm trial within a distributed network. This causal framework enables fair treatment comparisons across various target populations, addressing baseline imbalances and enhancing estimation robustness. Specifically, we estimate the expected counterfactual 40-month TBV changes for the four interventions and compare the effects of intensive BP versus standard BP, fibrate versus placebo and intensive BP versus fibrate across the BP trial population, lipid trial population, and overall population combining both trial populations. DAC estimates are compared with DOR estimates.

In our analysis, outcome models for 40-month TBV changes use cubic B-splines for continuous variables and include all covariates selected via forward selection in each therapy group.
These covariates include gender, history of cardiovascular disease, myocardial infarction, stroke, smoking status, education level, self-report history of depression, glycemia intervention, clinical center network, and spline functions of baseline TBV, intracranial volume, age, height and body mass index.  The same covariates are used for calibration weighting. The outcome models achieved R-squared values of 67.4\%, 66.8\%, 71.4\% and 76.5\% for the standard BP, intensive BP, lipid placebo, and lipid fibrate groups, respectively.

\begin{figure}[!ht]
\centering
\includegraphics[width=0.99\textwidth]{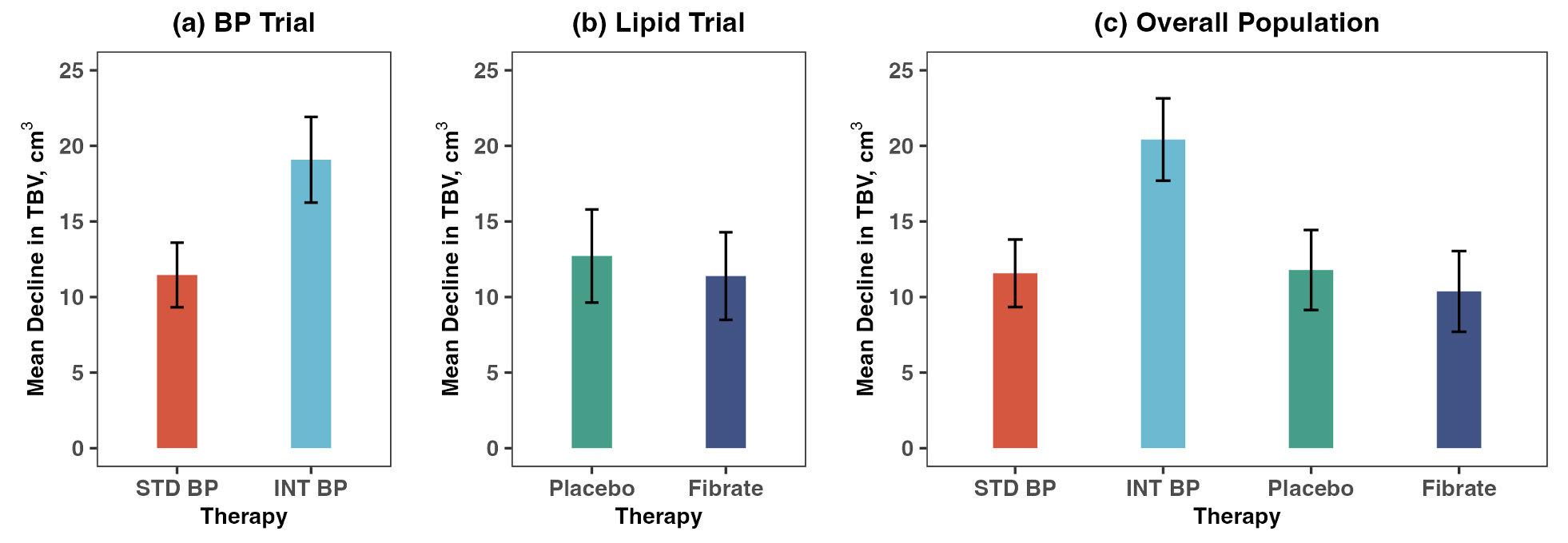}\\		
\caption{Means of counterfactual 40-month declines in total brain volume (TBV)  under four treatments: standard (STD) blood pressure, intensive (INT) blood pressure, lipid placebo  and lipid fibrate in (a) the blood pressure (BP) trial population,  (b) the lipid trial population and (c) the overall population combining BP and lipid trials.  Whiskers mark 95\% confidence intervals. \label{figure:decline}
}
\end{figure}

Figure \ref{figure:decline}  displays DAC estimates of counterfactual means of 40-month TBV declines under the four interventions in the BP trial, lipid trial, and overall populations, revealing significant TBV declines for all interventions. Estimates for the overall population differ slightly from trial-specific estimates due to heterogeneity between the BP and lipid trials, consistent with results in \cite{Williamson2014cognitive}. 
Table \ref{tb:ATE} summarizes the ATEs of 40-month TBV changes from baseline between different treatments. For intensive versus standard BP therapy, the DAC estimate in the overall population shows an 8.8 cm$^3$ [95\% CI, 5.5  to 12.2 cm$^3$; P-value = 2$\times 10^{-7}$] greater TBV decline under intensive therapy. 
No significant difference was observed between fibrate therapy and placebo. 
Both estimation methods yield consistent conclusions, aligning with findings of \cite{Williamson2014cognitive}. 
And DAC estimators exhibited slightly smaller variances than DOR estimators. These highlight the robustness and improved efficiency of DAC.

\begin{table}[ht]
\centering
\caption{The distribued outcome regression (DOR, Figure S3 in the SM) and distributed augmented calibration weighting (DAC, Figure \ref{fig:workflowDR}) estimators with their 95\% confidence intervals for average treatment effects (ATEs) on the  40-month changes of total brain volume  from baseline between standard (STD) blood pressure, intensive (INT) blood pressure, lipid placebo  and lipid fibrate treatments in the  blood pressure (BP) trial, the lipid trial, and  the overall populations. P-values from two-sided hypothesis tests  are listed in the last column. \label{tb:ATE}
}
\scalebox{0.86}{
\begin{tabular}{ccccc}
\hline
Target population  & Comparators & Method & ATE (95\% CI), cm$^3$ & P-value \\ 
\hline
BP trial population & INT BP v.s. STD BP 
& DOR & -7.7 (-11.3, -4.1) & $<0.001$ \\ 
&  & DAC & -7.6 (-10.9, -4.3) & $<0.001$ \\
\hline
Lipid trial population & Fibrate v.s. Placebo 
& DOR & 1.6 (-2.3, 5.5) & 0.41 \\ 
&  & DAC & 1.3 (-2.5, 5.2) & 0.50 \\ 
\hline
Overall population & INT BP v.s. STD BP 
& DOR & -9.2 (-13.1, -5.2) & $<0.001$ \\ 
&  & DAC & -8.8 (-12.2, -5.5) & $<0.001$ \\ 
\cline{2-5}
& Fibrate v.s. Placebo   
& DOR & 2.1 (-1.6, 5.8) & 0.27 \\ 
&  & DAC & 1.4 (-2.2, 5) & 0.44 \\ 
\cline{2-5}
& INT BP v.s. Fibrate 
& DOR & -11.1 (-15.5, -6.8) & $<0.001$ \\ 
&  &  DAC & -10.0 (-13.9, -6.2) & $<0.001$ \\ 
\hline
\end{tabular}}
\end{table}

\section{Conclusion}
\label{sec:conclusion} 
In this paper, we propose a novel collaborative indirect treatment comparison method to simultaneously estimate ATEs between any pair of treatments across various target populations in a distributed network of multiple single-arm trials without overlapping treatments. It introduces the DAC estimation, which is doubly robust consistent, accommodates data heterogeneity, and requires only aggregated data from each site. For parametric working models, the asymptotic variance of the DAC estimator depends on how nuisance parameters are estimated. To address this, we develop a bias-reduced DAC estimator in which nuisance parameters are estimated by adapting to the EIF, to achieve doubly robust inference and reduced first-order asymptotic bias. We provide the four-step, lossless DAC estimation procedure which requires two communication rounds between the coordinating center and participating sites. 

We extend the proposed estimation method to handle multi-arm trials with overlapping treatments across difference trials and consider the relaxation of the exchangeability assumption in Section S6 of the SM. A potential future direction is adapting the method for high-dimensional data, where numerous covariates are used for confounding adjustment via outcome regression and propensity score models. Approaches for doubly robust inference in high-dimensional settings have been developed for centralized frameworks \citep{Tan2020Model,Ning2020robust} and could be adapted for distributed networks. Another extension involves creating a debiased calibration estimator using generalized entropy \citep{kwon2024debiased} to improve estimation accuracy, especially in scenarios with complex relationships between the calibrated variables and outcomes.

\bibliographystyle{apalike}

\bibliography{cleanversion}

\begin{thebibliography}{}

\bibitem[Chan et~al., 2015]{Chan2015Globally}
Chan, K., Yam, S., and Zheng, Z. (2015).
\newblock Globally efficient non-parametric inference of average treatment
  effects by empirical balancing calibration weighting.
\newblock {\em Journal of the Royal Statistical Society Series B (Statistical
  Methodology)}, 78(3):673–700.

\bibitem[Chen, 2007]{CHEN2007Large}
Chen, X. (2007).
\newblock Large sample sieve estimation of semi-nonparametric models.
\newblock volume~6 of {\em Handbook of Econometrics}, page 5549–5632.
  Elsevier.

\bibitem[Chen and White, 1999]{Chen1999Improved}
Chen, X. and White, H. (1999).
\newblock Improved rates and asymptotic normality for nonparametric neural
  network estimators.
\newblock {\em IEEE Transactions on Information Theory}, 45(2):682--691.

\bibitem[Cheng et~al., 2020]{Cheng2020statistical}
Cheng, D., Ayyagari, R., and Signorovitch, J. (2020).
\newblock The statistical performance of matching-adjusted indirect
  comparisons: estimating treatment effects with aggregate external control
  data.
\newblock {\em The Annals of Applied Statistics}, 14(4):1806--1833.

\bibitem[Chernozhukov et~al., 2018]{chernozhukov2018double}
Chernozhukov, V., Chetverikov, D., Demirer, M., et~al. (2018).
\newblock Double/debiased machine learning for treatment and structural
  parameters.
\newblock {\em Econometrics Journal}, 21:C1–C68.

\bibitem[Dagenais et~al., 2022]{dagenais2022use}
Dagenais, S., Russo, L., Madsen, A., et~al. (2022).
\newblock Use of real-world evidence to drive drug development strategy and
  inform clinical trial design.
\newblock {\em Clinical Pharmacology \& Therapeutics}, 111(1):77--89.

\bibitem[Deville and Särndal, 1992]{Deville1992Calibration}
Deville, J.-C. and Särndal, C.-E. (1992).
\newblock Calibration estimators in survey sampling.
\newblock {\em Journal of the American Statistical Association},
  87(418):376--382.

\bibitem[{European Commission}, 2024a]{eu2024jca1}
{European Commission} (2024a).
\newblock {Guidance on outcomes for joint clinical assessments}.

\bibitem[{European Commission}, 2024b]{eu2024jca2}
{European Commission} (2024b).
\newblock {Guidance on the validity of clinical studies for joint clinical
  assessments}.

\bibitem[{European Union}, 2021]{eu2021hta}
{European Union} (2021).
\newblock {Regulation (EU) 2021/2282 of the European Parliament and of the
  Council of 15 December 2021 on Health Technology Assessment}.

\bibitem[{Food and Drug Administration}, 2001]{FDA2001guidance}
{Food and Drug Administration} (2001).
\newblock Guidance for industry: E10 choice of control group and related issues
  in clinical trials.
\newblock \url{https://www.fda.gov/media/71349/}.
\newblock Accessed September 25, 2024.

\bibitem[Funk et~al., 2011]{funk2011doubly}
Funk, M.~J., Westreich, D., Wiesen, C., et~al. (2011).
\newblock Doubly robust estimation of causal effects.
\newblock {\em American Journal of Epidemiology}, 173(7):761--767.

\bibitem[Gray et~al., 2020]{gray2020framework}
Gray, C.~M., Grimson, F., Layton, D., et~al. (2020).
\newblock A framework for methodological choice and evidence assessment for
  studies using external comparators from real-world data.
\newblock {\em Drug Safety}, 43:623--633.

\bibitem[Gu and Chen, 2023]{gu2023distributed}
Gu, J. and Chen, S.~X. (2023).
\newblock Distributed statistical inference under heterogeneity.
\newblock {\em Journal of Machine Learning Research}, 24(387):1--57.

\bibitem[Guo et~al., 2025]{Guo2025Robust}
Guo, Z., Li, X., Han, L., et~al. (2025).
\newblock Robust inference for federated meta-learning.
\newblock {\em Journal of the American Statistical Association}, 0(0):1--16.

\bibitem[Hainmueller, 2012]{Hainmueller2012entropy}
Hainmueller, J. (2012).
\newblock Entropy balancing for causal effects: A multivariate reweighting
  method to produce balanced samples in observational studies.
\newblock {\em Political analysis}, 20(1):25--46.

\bibitem[Han et~al., 2025]{Han2025Federated}
Han, L., Hou, J., Cho, K., et~al. (2025).
\newblock {Federated Adaptive Causal Estimation (FACE)} of target treatment
  effects.
\newblock {\em Journal of the American Statistical Association}.

\bibitem[Hatswell et~al., 2016]{hatswell2016regulatory}
Hatswell, A.~J., Baio, G., Berlin, J.~A., et~al. (2016).
\newblock Regulatory approval of pharmaceuticals without a randomised
  controlled study: {A}nalysis of {EMA and FDA} approvals 1999--2014.
\newblock {\em BMJ Open}, 6(6):e011666.

\bibitem[Hu et~al., 2024]{Hu2024Collaborative}
Hu, M., Shi, X., and Song, P. X.-K. (2024).
\newblock Collaborative inference for treatment effect with distributed
  data-sharing management in multicenter studies.
\newblock {\em Statistics in Medicine}, 43(11):2263--2279.

\bibitem[Jahanshahi et~al., 2021]{jahanshahi2021use}
Jahanshahi, M., Gregg, K., Davis, G., et~al. (2021).
\newblock The use of external controls in fda regulatory decision making.
\newblock {\em Therapeutic Innovation \& Regulatory Science}, 55(5):1019--1035.

\bibitem[Kim and Haziza, 2014]{Kim2014doubly}
Kim, J.~K. and Haziza, D. (2014).
\newblock Doubly robust inference with missing data in survey sampling.
\newblock {\em Statistica Sinica}, 24(1):375--394.

\bibitem[Kwon et~al., 2024]{kwon2024debiased}
Kwon, Y., Kim, J.~K., and Qiu, Y. (2024).
\newblock Debiased calibration estimation using generalized entropy in survey
  sampling.
\newblock {\em arXiv}.

\bibitem[Lee, 2018]{Lee2018Efficient}
Lee, Y.-Y. (2018).
\newblock Efficient propensity score regression estimators of multivalued
  treatment effects for the treated.
\newblock {\em Journal of Econometrics}, 204(2):207--222.

\bibitem[Li and Song, 2020]{Li2020Target}
Li, X. and Song, Y. (2020).
\newblock Target population statistical inference with data integration across
  multiple sources — an approach to mitigate information shortage in rare
  disease clinical trials.
\newblock {\em Statistics in Biopharmaceutical Research}, 12(3):322--333.

\bibitem[Lunceford and Davidian, 2004]{Lunceford2004strat}
Lunceford, J.~K. and Davidian, M. (2004).
\newblock Stratification and weighting via the propensity score in estimation
  of causal treatment effects: a comparative study.
\newblock {\em Statistics in Medicine}, 23(19):2937--2960.

\bibitem[Miksad et~al., 2019]{miksad2019small}
Miksad, R.~A., Samant, M.~K., Sarkar, S., et~al. (2019).
\newblock Small but mighty: the use of real-world evidence to inform precision
  medicine.
\newblock {\em Clinical Pharmacology \& Therapeutics}, 106(1):87--90.

\bibitem[Ning et~al., 2020]{Ning2020robust}
Ning, Y., Sida, P., and Imai, K. (2020).
\newblock Robust estimation of causal effects via a high-dimensional covariate
  balancing propensity score.
\newblock {\em Biometrika}, 107(3):533--554.

\bibitem[Patel et~al., 2021]{patel2021use}
Patel, D., Grimson, F., Mihaylova, E., et~al. (2021).
\newblock Use of external comparators for health technology assessment
  submissions based on single-arm trials.
\newblock {\em Value in Health}, 24(8):1118--1125.

\bibitem[Phillippo et~al., 2016]{phillippo2016nice}
Phillippo, D., Ades, T., Dias, S., et~al. (2016).
\newblock {NICE DSU technical support document 18: Methods for
  population-adjusted indirect comparisons in submissions to NICE}.
\newblock
  {https://research-information.bris.ac.uk/ws/portalfiles/portal/94868463}.
\newblock Accessed September 25, 2024.

\bibitem[Phillippo et~al., 2018]{phillippo2018methods}
Phillippo, D.~M., Ades, A.~E., Dias, S., et~al. (2018).
\newblock Methods for population-adjusted indirect comparisons in health
  technology appraisal.
\newblock {\em Medical Decision Making}, 38(2):200--211.

\bibitem[Qin and Zhang, 2007]{Qin2007empirical}
Qin, J. and Zhang, B. (2007).
\newblock Empirical-likelihood-based inference in missing response problems and
  its application in observational studies.
\newblock {\em Journal of the Royal Statistical Society Series B: Statistical
  Methodology}, 69(1):101--122.

\bibitem[Rubin, 1974]{rubin1974estimating}
Rubin, D.~B. (1974).
\newblock Estimating causal effects of treatments in randomized and
  nonrandomized studies.
\newblock {\em Journal of Educational Psychology}, 66(5):688.

\bibitem[Schmidli et~al., 2014]{schmidli2014robust}
Schmidli, H., Gsteiger, S., Roychoudhury, S., et~al. (2014).
\newblock Robust meta-analytic-predictive priors in clinical trials with
  historical control information.
\newblock {\em Biometrics}, 70(4):1023--1032.

\bibitem[Seymour et~al., 2010]{seymour2010design}
Seymour, L., Ivy, S.~P., Sargent, D., et~al. (2010).
\newblock The design of phase {II} clinical trials testing cancer therapeutics:
  {C}onsensus recommendations from the clinical trial design task force of the
  national cancer institute investigational drug steering committee.
\newblock {\em Clinical Cancer Research}, 16(6):1764--1769.

\bibitem[Shi et~al., 2023]{shi2023data}
Shi, X., Pan, Z., and Miao, W. (2023).
\newblock Data integration in causal inference.
\newblock {\em Wiley Interdisciplinary Reviews: Computational Statistics},
  15(1):e1581.

\bibitem[Signorovitch et~al., 2010]{Signorovitch2010Comparative}
Signorovitch, J., Wu, E., Yu, A., et~al. (2010).
\newblock Comparative effectiveness without head-to-head trials a method for
  matching-adjusted indirect comparisons applied to psoriasis treatment with
  adalimumab or etanercept.
\newblock {\em PharmacoEconomics}, 28(10):935--945.

\bibitem[Stuart et~al., 2011]{Stuart2011Use}
Stuart, E., Cole, S., Bradshaw, C., et~al. (2011).
\newblock The use of propensity scores to assess the generalizability of
  results from randomized trials.
\newblock {\em Journal of the Royal Statistical Society Series A},
  174(2):369--386.

\bibitem[Tan, 2020]{Tan2020Model}
Tan, Z. (2020).
\newblock Model-assisted inference for treatment effects using regularized
  calibrated estimation with high-dimensional data.
\newblock {\em Annals of Statistics}, 48(2):811–837.

\bibitem[Thorlund et~al., 2020]{thorlund2020synthetic}
Thorlund, K., Dron, L., Park, J.~J., et~al. (2020).
\newblock Synthetic and external controls in clinical trials--a primer for
  researchers.
\newblock {\em Clinical epidemiology}, 12:457--467.

\bibitem[{Van der Vaart}, 2002]{vander2002}
{Van der Vaart}, A. (2002).
\newblock {\em Semiparametric statistics}, pages 331--457.
\newblock Number 1781 in Lecture Notes in Math. Springer.

\bibitem[Vermeulen and Vansteelandt, 2015]{vermeulen2015bias}
Vermeulen, K. and Vansteelandt, S. (2015).
\newblock Bias-reduced doubly robust estimation.
\newblock {\em Journal of the American Statistical Association},
  110(511):1024--1036.

\bibitem[Vickers et~al., 2007]{vickers2007setting}
Vickers, A.~J., Ballen, V., and Scher, H.~I. (2007).
\newblock Setting the bar in phase {II} trials: {T}he use of historical data
  for determining “go/no go” decision for definitive phase {III} testing.
\newblock {\em Clinical Cancer Research}, 13(3):972--976.

\bibitem[Wager and Athey, 2018]{Wager2018Estimation}
Wager, S. and Athey, S. (2018).
\newblock Estimation and inference of heterogeneous treatment effects using
  random forests.
\newblock {\em Journal of the American Statistical Association},
  113(523):1228--1242.

\bibitem[Wang et~al., 2024]{Wang2024comparison}
Wang, H., Wu, F., and Chen, Y.-F. (2024).
\newblock A comparison of methods for estimating the average treatment effect
  on the treated for externally controlled trials.
\newblock {\em arXiv}.

\bibitem[Weber et~al., 2018]{weber2018predicting}
Weber, K., Hemmings, R., and Koch, A. (2018).
\newblock How to use prior knowledge and still give new data a chance?
\newblock {\em Pharmaceutical Statistics}, 17(4):329--341.

\bibitem[Williams and Evans, 2023]{william2023the}
Williams, D.~M. and Evans, M. (2023).
\newblock The evolution of real-world evidence in healthcare decision making.
\newblock {\em Expert Opinion on Drug Safety}, 22(6):443--445.

\bibitem[Williamson et~al., 2014]{Williamson2014cognitive}
Williamson, J.~D., Launer, L.~J., Bryan, R.~N., et~al. (2014).
\newblock Cognitive function and brain structure in persons with type 2
  diabetes mellitus after intensive lowering of blood pressure and lipid
  levels: a randomized clinical trial.
\newblock {\em JAMA internal medicine}, 174(3):324--333.

\bibitem[Woolacott et~al., 2017]{woolacott2017methodological}
Woolacott, N., Corbett, M., Jones-Diette, J., et~al. (2017).
\newblock Methodological challenges for the evaluation of clinical
  effectiveness in the context of accelerated regulatory approval: {A}n
  overview.
\newblock {\em Journal of Clinical Epidemiology}, 90:108--118.

\bibitem[Xiong et~al., 2023]{Xiong2023Federated}
Xiong, R., Koenecke, A., Powell, M., et~al. (2023).
\newblock Federated causal inference in heterogeneous observational data.
\newblock {\em Statistics in Medicine}, 42(24):4418--4439.

\bibitem[Zhao and Percival, 2016]{Zhao2016Entropy}
Zhao, Q. and Percival, D. (2016).
\newblock Entropy balancing is doubly robust.
\newblock {\em Journal of Causal Inference}, 5(1):20160010.

\bibitem[Zubizarreta, 2015]{Zubizarreta2015stable}
Zubizarreta, J.~R. (2015).
\newblock Stable weights that balance covariates for estimation with incomplete
  outcome data.
\newblock {\em Journal of the American Statistical Association},
  110(511):910--922.

\end{thebibliography}

\end{document}